# Membrane-Electrode Assemblies for Electrochemical Reduction of $CO_2$ to Ethylene: Design for Minimal Energy Consumption

Tugrul Y. Ertugrul*, Woong Choi, Adam Z. Weber, Alexis T. Bell*

Tugrul Y. Ertugrul
Energy Technologies Area
Lawrence Berkeley National Laboratory
Berkeley, CA 94720, USA

Current address:
Henry Samueli School of Engineering
University of California Los Angeles
Los Angeles, CA 90095, USA
E-mail: tyertugrul@ucla.com

Woong Choi
Energy Technologies Area
Lawrence Berkeley National Laboratory
Berkeley, CA 94720, USA

Current address:
Departments of Energy Engineering and Energy Systems Engineering
Gyeongsang National University
Jinju-si, Gyeongnam 52849, Republic of Korea

Adam Z. Weber
Energy Technologies Area
Lawrence Berkeley National Laboratory
Berkeley, CA 94720, USA

Alexis T. Bell
Department of Chemical and Biomolecular Engineering
University of California Berkeley
Berkeley, CA 94720, USA
E-mail: alexbell@berkeley.edu

Supporting information for this article is given via a link at the end of the document.

# Abstract

Membrane-electrode-assembly (MEA) cells with copper (Cu) cathodes show strong potential for electrochemical $CO_2$ reduction to ethylene ($C_2H_4$), but achieving high $C_2H_4$ selectivity remains a challenge due to competing hydrogen evolution. This selectivity is highly sensitive to the local microenvironment near the Cu catalyst surface. In this study, a 1-D, multiphysics continuum model is utilized to investigate how MEA cell performance and faradaic efficiency (FE) to $C_2H_4$ are affected by both component properties and operating conditions, with particular focus on coupled transport and reaction phenomena. Key parameters include cathode electrochemically active surface area (ECSA) and catalyst layer thickness. Halving catalyst layer thickness increases FE to $C_2H_4$ by 2% and lowers the cell voltage by 40 mV. In contrast, a tenfold decrease in ECSA results increases the FE to $C_2H_4$ by 7% but leads increase cell voltage at a given current density by 150 mV. This tradeoff occurs because the potential distribution within the cathode catalyst layer is the primary driving force for $C_2H_4$ formation. Increased cell voltage also raises the energy cost of $C_2H_4$ production. This model framework enables techno-economic assessments and identifies key factors that must be optimized to enable economically viable production of $C_2H_4$ via electrochemical reduction of $CO_2$.

## Graphical Abstract

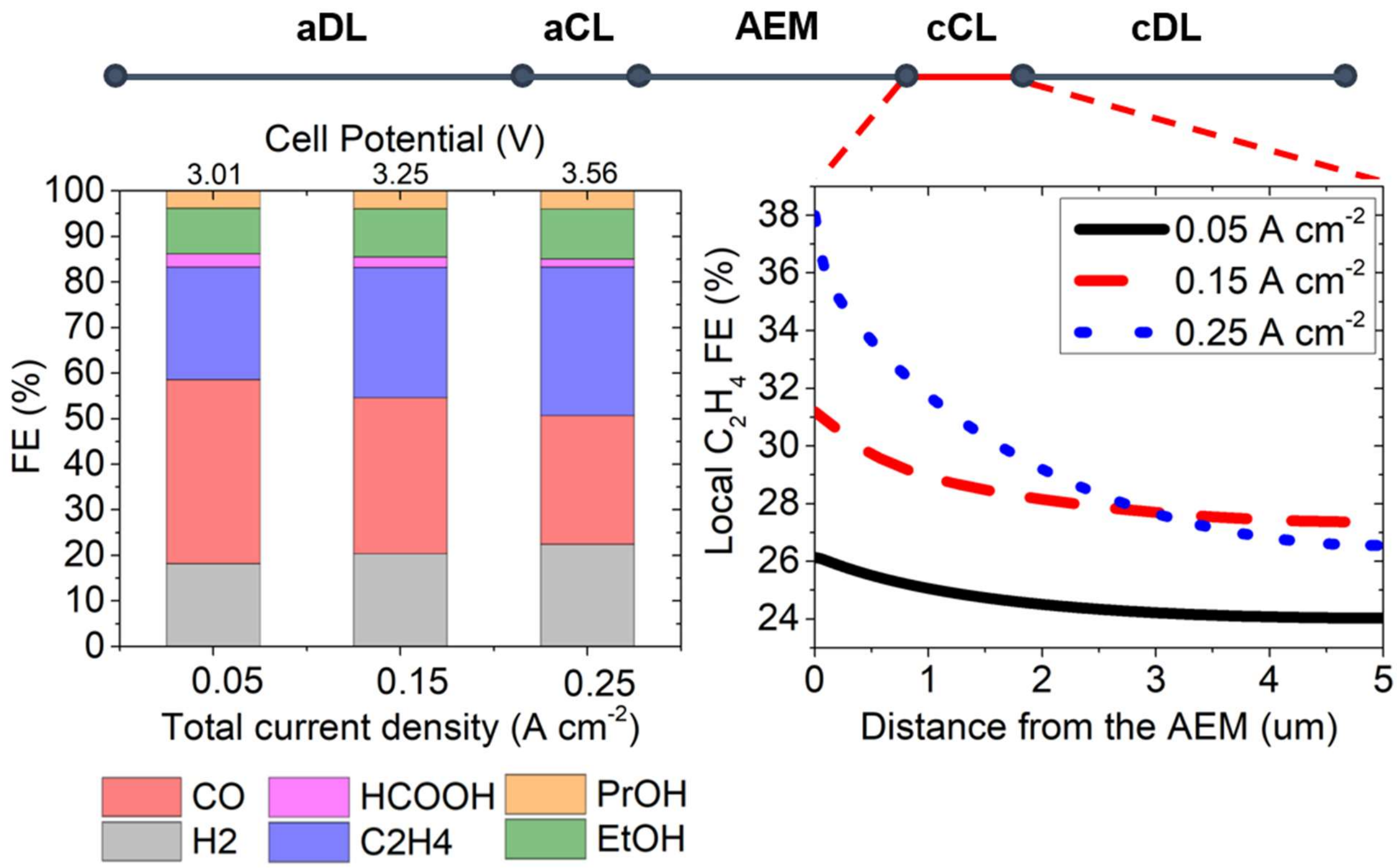

# 1. Introduction

Electrochemical $CO_2$ reduction ($CO_2R$) provides an option for producing fuels and chemical feedstocks from $CO_2$ emitted from stationary sources using electricity [1–4]. One of the most attractive products of $CO_2R$ is ethylene ($C_2H_4$), because of its large market size (180 Mt $y^{-1}$) [5,6] and viability for production via direct electrochemical $CO_2$ reduction. Multiple studies have shown that copper (Cu) is the most effective catalyst for the electrochemical reduction of $CO_2$ to $C_2H_4$ [7,8]; however, Cu can also produce other products, e.g., hydrogen ($H_2$), carbon monoxide (CO), formic acid (HCOOH), methane ($CH_4$), ethanol ($C_2H_5OH$), and propanol ($C_3H_7OH$) [8–14].

A recent techno-economic analysis suggests that for the industrial production of $C_2H_4$, the faradaic efficiency (FE) for $C_2H_4$ formation should exceed 60% and the total current should exceed 0.3 A $cm^{-2}$ [15–17]. These targets can be achieved using gas-diffusion electrodes (GDEs) incorporated into a zero-gap membrane-electrode assembly (MEA) [18–20]. It is also known that the activity and FE to $C_2H_4$ are very sensitive to the microenvironment near the catalyst surface [13,21,22]. Given the complexity of the multiple transport and reaction phenomena that affect MEA performance, theoretical modeling is ideally suited to explore how $C_2H_4$ partial current density and FE depend on the properties of the components comprising the MEA, as well as its operating conditions [21–26]. Once validated against experimental observations, such a model can be used to define the fractional utilization of $CO_2$ and the energy required to produce a kilogram of $C_2H_4$, information required to establish the economic viability of producing $C_2H_4$ via $CO_2R$ in an MEA.

Several efforts aimed at stimulating the performance of MEAs for $CO_2R$ have been reported [24,27–29]. The earliest comprehensive full-MEA modeling study focused on simulating $CO_2R$ to CO and $H_2$ on an Ag catalyst, using a one-dimensional (1-D), multiphysics model [29]. While the computational domain included an anion-exchange membrane (AEM), a cathode catalyst layer, and a cathode diffusion layer (cDL), the entire anode was treated as an interface. Each layer was characterized by the volume fraction of gas, liquid, and solid phases, and the transport and reactions (bulk and electrochemical) of ionic and neutral components within each phase and between phases was considered explicitly. The kinetics of homogeneous buffer reactions (*i.e.*, reactions involving the interconversion of $OH^-$, $H^+$, $CO_2$, $HCO_3^-$, and $CO_3^{2-}$) occurring in the liquid phase were treated explicitly and all electrochemical reactions were described by Tafel kinetics. This study also considered the effects of water content in the ionomer present in the cathode catalyst layer and in the AEM on the overall performance of the MEA for $CO_2R$, *i.e.*, the total current density and the product selectivity between $H_2$ and CO. Subsequent work extended this 1-D approach to $CO_2R$ over Cu with the aim of describing the effects of cell voltage on the distribution of products ($H_2$, CO, HCOOH, $C_2H_4$, $C_2H_5OH$, $C_3H_7OH$) formed. In this model, an aDL and an anode catalyst layer were also included in the computational domain [30]. Similar 1-D models have been employed to investigate carbonate crossover which affects $CO_2$ utilization during $CO_2R$ in an MEA [31,32]. Other 1-D models have been utilized to examine the importance of water transport for electrochemical $CO_2R$ in an MEA [33,34]. 2-D models of $CO_2R$ in an MEA have also been reported to account for spatial variations along the flow channel [35,36]. A recent 2-D MEA model performed a sensitivity analysis on a wide range of parameters affecting $CO_2R$ performance on an Ag catalyst [37]. While the studies cited above have been used to explore the effects of the parameters affecting the performance of $CO_2R$ in MEAs (*e.g.*, membrane thickness, cathode catalyst-layer thickness, hydration of the ionomer and membrane, *etc.*), to the best of our knowledge, a comprehensive study of all factors affecting to $C_2H_4$ production over Cu occurring in an MEA has not been reported previously.

This study investigates strategies for reducing the voltage required to achieve a given overall current density and increase the FE for $C_2H_4$ in a Cu-based $CO_2$R MEA cell. To this end, a 1-D, steady-state, continuum model, validated against experimental measurements, was developed. This model was then used to assess the effects of multiple design and operating parameters, (*i.e.*, $CO_2$ partial pressure, membrane thickness, anode and cathode catalyst-layer thickness, anode and cathode electrochemically-active surface area (ECSA), and ion and water transport through the membrane) on the energy required to produce a kilogram of $C_2H_4$.

# 2. Methods

A 1-D, steady-state continuum model was developed using COMSOL Multiphysics® software based on our previous $CO_2$R electrolyzer models [28–30]. The computational domain consists of an AEM at the center, anode/cathode catalyst layers (aCL/cCL), and anode and cathode DLs (aDL/cDL), as shown in Figure 1. Governing equations and continuum relationships including conservation of mass, momentum, species, and charge coupled with Butler-Volmer kinetics are discussed in the following sections.

## 2.1 Reaction Kinetics

Acidic and alkaline oxygen-evolution reactions (OER) occur on an iridium-oxide catalyst at the anode:

$$2\mathrm{H_2O} \rightarrow 4\mathrm{e^-} + \mathrm{O_2} + 4\mathrm{H^+}, \qquad U_{OER}^{0} = 1.23\, V\, vs.SHE, pH = 0 \tag{1}$$

$$4\mathrm{OH^-} \rightarrow 4\mathrm{e^-} + \mathrm{O_2} + 2\mathrm{H_2O}, \qquad U_{OER}^{0} = 1.23\, V\, vs.SHE, pH = 0 \tag{2}$$

At the cathode, copper catalyzes hydrogen-evolution reaction (HER) and $CO_2$R reactions: carbon-monoxide (CO), formic acid (HCOOH), ethylene ($C_2H_4$), ethanol ($C_2H_5OH$), and propanol ($C_3H_7OH$),

$$2H_2O + 2e^- \rightarrow H_2 + 2OH^-, \qquad U^0_{HER} = 0\ V\ vs. SHE \tag{3}$$

$$CO_2 + H_2O + 2e^- \rightarrow CO + 2OH^-, \qquad U^0_{COER} = -0.06\ V\ vs. SHE \tag{4}$$

$$CO_2 + H_2O + 2e^- \rightarrow HCOO^- + 2OH^-, \qquad U^0_{HCOO^-} = -0.07\ V\ vs. SHE \tag{5}$$

$$2CO_2 + 8H_2O + 12e^- \rightarrow C_2H_4 + 12OH^-, \qquad U^0_{C_2H_4} = 0.09\ V\ vs. SHE \tag{6}$$

$$2CO_2 + 9H_2O + 12e^- \rightarrow C_2H_5OH + 12OH^-, \qquad U^0_{C_2H_5OH} = 0.10\ V\ vs. SHE \tag{7}$$

$$3CO_2 + 13H_2O + 18e^- \rightarrow C_3H_7OH + 18OH^-, U^0_{C_3H_7OH} = 0.11\ V\ vs. SHE \tag{8}$$

The standard thermodynamic potentials for each product were adopted from our previous work [22]. The partial current density of each product $k$ was represented by the concentration-dependent Tafel equation:

$$i_k = i_{0,k} \prod_j \left(\frac{c_{j,k}}{c_{j,k}^{ref}}\right)^{\gamma_{j,k}} exp\left(\frac{\alpha_{a/c,k}F}{RT}\eta_k\right) \tag{9}$$

where $i_{0,k}$ is the exchange current density, $\alpha_{c,k}$ is the charge transfer coefficient, $F$ is Faraday's constant (96485 C mol$^{-1}$), $R$ is the ideal-gas constant (8.314 J mol$^{-1}$ K$^{-1}$), $T$ is the absolute temperature. The overpotential $\eta_k$ was calculated by

$$\eta_k = \phi_s - \phi_l - \left(U^0_k - \frac{2.303RT}{F}pH\right) \tag{10}$$

where $\phi_s$ is the solid-phase potential (electric potential), $\phi_l$ is the liquid-phase potential, and $U^0_k$ is the standard thermodynamic potential. The expression $\frac{2.303RT}{F}$pH represents the Nernstian correction due to local pH change [38]. The concentration term $\left(\frac{c_j}{c_j^{ref}}\right)^{\gamma_{j,k}}$ considers $OH^-$ and $CO_{2(aq)}$ and water activities in the catalyst layers, where $c_j$ is the local concentration, $c_j^{ref}$ is the reference concentration and $\gamma_{j,k}$ is the activity coefficient. The reference concentrations were $c^{ref}_{OH^-} = 1\text{M}$ and $c^{\text{ref}}_{\text{CO}_{2(\text{aq})}} = 0.034\text{M}$, where the latter is the maximum solubility limit for the $CO_{2(aq)}$ in water. The water activity $a_w$ is defined in eqn. (35). All parameters are given in Table S1 in the SI.

The volumetric source term due to the porous-electrode reactions is

$$R_k = -\sum_k a_{s,k} \frac{\nu_{j,k} i_k}{n_k F} \tag{11}$$

where $n_k$ is the number of electrons transferred in reaction $k$, $\nu_{j,k}$ is the stoichiometric coefficient of species $j$ for reaction $k$, and $a_{s,k}$ is the electrochemically active surface area (ECSA), which is corrected for the catalyst-layer saturation $S_{CL}$ (the fraction of pores in the catalyst layer filled with water) as

$$a_{s,k} = (1 - S_{CL}) a_s^0 \tag{12}$$

The capillary pressure ($P_c = P_L - P_G$) is used to determine $S_{CL}$ and its relationship is shown in Figure S1a in the SI [39]. It is assumed that $CO_2$ reduction primarily occurs at the catalyst surface exposed to $CO_2$ gas, while it is significantly limited at surfaces covered by the electrolyte due to the low solubility of $CO_2$ in the electrolyte [30]. The CL specific-surface area $a_s^0$ is determined using geometric relationship [28]

$$a_s^0 = \frac{3\varepsilon_s}{r_p} \tag{13}$$

where $r_p$ is the average catalyst particle, $\varepsilon_s$ is the solid phase volume fraction of the catalyst layer defined as

$$\varepsilon_s = 1 - \varepsilon_{CL}^0 \tag{14}$$

where $\varepsilon_{CL}^{0}$ is the porosity of the catalyst layer, listed in Table S3 in the SI.

## 2.2 Homogeneous Bulk Reactions

The following homogeneous bulk reactions were considered in the ionomer and the membrane:

$$CO_2(aq) + H_2O \xleftrightarrow{k_1, k_{-1}} H^+ + HCO_3^- \qquad K_1 \tag{15}$$

$$HCO_3^- \xleftrightarrow{k_2, k_{-2}} H^+ + CO_3^{2-} \qquad K_2 \tag{16}$$

$$CO_2(aq) + OH^- \xleftrightarrow{k_3, k_{-3}} HCO_3^- \qquad K_3 \tag{17}$$

$$HCO_3^- + OH^- \xleftrightarrow{k_4, k_{-4}} H_2O + CO_3^{2-} \qquad K_4 \tag{18}$$

$$H_2O \xleftrightarrow{k_5, k_{-5}} H^+ + OH^- \qquad K_5 \tag{19}$$

$$HCOOH \xleftrightarrow{k_6, k_{-6}} H^+ + HCOO^- \qquad K_6 \tag{20}$$

where $K_n$ is the equilibrium constant and $k_n$ and $k_{-n}$ are the rate constants for the forward and reverse directions of reaction $n$, respectively [22]. The volumetric molar source term is defined as

$$R_{B,j} = \sum_j \nu_{j,n} \left( k_n \prod_{\nu_{j,n} < 0} {c_j}^{-\nu_{j,n}} - \frac{k_n}{K_n} \prod_{\nu_{j,n} > 0} {c_j}^{\nu_{j,n}} \right) \tag{21}$$

where $\nu_{j,n}$ is the sociocentric coefficient of species $j$ and the bulk reaction $n$ and $c_j$ is concentration of species $j$. The sign is negative for reactants and positive for products. All homogeneous bulk reactions and parameters are listed in Table S2 in the SI.

## 2.3 Charge Conservation and Transport

Electronic current conduction (electron transport) in the solid phase (CLs and DLs) was described by Ohm's law:

$$i_s = -\sigma_{s,m}^{eff} \nabla\phi_s \tag{22}$$

where $i_s$ is the solid phase current density, $\phi_s$ is the solid-phase potential, $\sigma_{s,m}^{eff}$ is the solid phase effective conductivity of porous medium $m$, calculated using Bruggeman correlation,

$$\sigma_{s,m}^{eff} = (\varepsilon_{s,m})^{1.5}\sigma_{s,m} \tag{23}$$

where $\varepsilon_{s,m}$ is the solid volume fraction and $\sigma_{s,m}$ is the intrinsic electronic conductivity of the porous medium $m$, tabulated in Table S3 in the SI.

We considered the transport of key ionic species including $OH^-$, $H^+$ $HCO_3^-$, $CO_3^{2-}$, and $HCOO^-$ within the AEM and the CLs. The transport of these ions was modeled using the Nernst-Plank equation, accounting diffusion, migration, and convection. Due to the selective nature of the AEM, cation transport (e.g., $K^+$) was assumed to be minimal and was therefore not explicitly simulated.

$$\mathbf{N}_j = -D_j^{\mathrm{eff}}\nabla c_j - z_j u_{mob,j} F c_j \nabla\phi_l + c_j \mathbf{u}_L \tag{24}$$

where $c_j$ is the concentration of species $j$, $z_j$ is the charge number of species $j$, $\phi_I$ liquid-phase potential, $u_{mob,j}$ is the mobility of species $j$, $\mathbf{u}_L$ is the liquid mass-averaged velocity calculated by Darcy's law. The effective diffusivity is given by

$$D_j^{eff} = (f_{I,CL}\varepsilon_{CL}^0)^{1.5}\mathrm{D}_j \tag{25}$$

where $f_{I,CL}$ is the ionomer volume fraction in CL pore space (provided in Table S1 in the SI), $D_j$ is the nominal diffusivity of species $j$. The diffusion coefficient of ionic species in the membrane/ionomer are calculated from the membrane/ionomer conductivity correlation as

$$D_j = \frac{K_j^{eff,M/I} RT}{z_j F^2} \tag{26}$$

where $K^{eff,M/I}$ is the effective membrane/ionomer conductivity, depended on the saturation of the membrane/ionomer,

$$K^{eff,M/I} = S_m K_L + (1 - S_m) K_V \tag{27}$$

where $K_L$ is liquid equilibrated conductivity, $K_V$ is the vapor equilibrated conductivity, listed in Table S4 in the SI. The membrane/ionomer saturation $S_m$ is determined from the $P_{L,M}$ . The relationship between $P_{L,M}$ and $S_m$ is shown in Figure S1c in the SI [40]. The conductivity values were adopted from the literature for the $HCO_3^-$ form AEM [30,41–43]; It is assumed that the $CO_3^-$ and $HCOO^-$form AEM have the same conductivity, while the $H^+$form AEM has a $10 \times$ lower conductivity and the $OH^-$ form AEM has a $5 \times$ higher conductivity [30].

The mobility was determined by the Nernst-Einstein relation:

$$u_{mob,j} = \frac{D_j}{RT} \tag{28}$$

The electrolyte current density was calculated using Faraday's law by summing up the contribution from the molar fluxes of all ionic species,

$$\mathbf{i}_l = F \sum_j z_j \left( -D_j^{eff} \nabla c_j - z_j u_{mob,j} F c_j \nabla \phi_l \right) \tag{29}$$

where the convection term vanishes due to the assumption of electroneutrality,

$$\sum_j z_j c_j = 0 \tag{30}$$

The conservation of charge is then used to relate the solid-phase electric potential, $\phi_s$ and liquid-phase potential $\phi_l$,

$$\nabla \cdot i_s = -\nabla \cdot i_l = -F\sum_k z_k R_k - \varepsilon_l F\sum_j z_j R_{B,j} \tag{31}$$

## 2.4 Water Transport in The Membrane/ionomer Phase

The molar flux of water in the membrane/ionomer phase (AEM and CLs) was calculated by

$$\mathbf{N}_w = -\alpha_w^{eff,M}\nabla\mu_w + \sum_j \zeta_j^{eff,M} N_j \tag{32}$$

where $\alpha_w^{eff,M}$ is the water transport coefficient and $\zeta_j^{eff,M}$ is the effective electro-osmotic coefficient of species $j$. Nominal values for the water transport coefficient and electro-osmotic coefficient are listed in Table S4 in the SI. Similar to the membrane conductivity, these values are adjusted based on membrane saturation, $S_m$. An overall apparent electro-osmotic coefficient can be defined as

$$\sum_j \zeta_j^{eff,M} N_j = -\zeta_j^{eff,M}\frac{i_l}{F} \tag{33}$$

The chemical potential of water, $\mu_w$, is defined as

$$\mu_w = RT\ln(a_w) + V_{m,w}(P_{L,M} - P_L^{ref}) \tag{34}$$

where $a_w$ is the activity of water, $V_{\mathrm{m},w}$ is the molar volume of liquid water (18 mL mol$^{-1}$), $p_{L,M}$ is the pressure of liquid water in the membrane/ionomer phase, and $p_{L,ref}$ is the reference liquid pressure (101 kPa). The conservation of water in the membrane/ionomer can be written as

$$\nabla \cdot N_w = R_{j,w} \tag{35}$$

where the source term includes water involved in electrochemical reactions and homogeneous buffer reactions (based on the eqn. (11)), as well as phase transfer of both vapor water and liquid water in the ionomer:

$$R_{PT,w,I} = a_s k_{MT,V}\left(\frac{RH}{100} - a_w\right) + \frac{a_s k_{MT,L}}{RT}\left(P_L - P_{L,M}\right) \tag{36}$$

where $k_{MT,V}$ is the mass-transfer coefficient for vapor water and $k_{MT,L}$ is the mass-transfer coefficient for liquid water set to $0.06\ mol\ m^{-2}s^{-1}$ and $10^4\ m\ s^{-1}$ respectively [41]. The relative humidity of the gas phase is $RH = \frac{P_G y_0}{P_0^{sat}} \times 100\%$.

## 2.5 Conservation Equations for Liquid and Gas Mixtures

Liquid and gas mixture ($CO_2$,$H_2$,$O_2$,$CO$,$C_2H_4$,$N_2$, and $H_2O$) transport in the catalyst layers and porous transport layers were modeled by Darcy's law,

$$u_p = \frac{K_{m,p}^{eff}}{\mu_p} \nabla p_p \tag{37}$$

$$K_{m,p}^{eff} = \left(\varepsilon_{m,p}\right)^{1.5} K_{m,p} \tag{38}$$

where $u_p$, $\mu_p$, and $p_p$ are the velocity, viscosity, and pressure of phase $p$, respectively. $K_{m,p}^{eff}$ is the effective permeability, $K_{m,p}$ is the intrinsic permeability, $\varepsilon_{m,p}$ is the volume fraction of medium $m$, and phase $p$. The volume fractions for liquid and gas phases in medium $m$ are calculated as

$$\varepsilon_{m,L} = \varepsilon_m^0(1 - f_{I,m})S_m \tag{39}$$

$$\varepsilon_{m,G} = \varepsilon_m^0(1 - f_{I,m})(1 - S_m) \tag{40}$$

It should be noted that the $f_{I,m}$ is zero for DM, as there is no ionomer present. All intrinsic properties of CL and DL are tabulated in Table S3 in the SI. Conservation of mass for both liquid and gas phases were solved to calculate pressure and velocity distributions,

$$\nabla \cdot (\rho_p u_p) = Q_p \tag{41}$$

where $\rho_p$ is the density, and $Q_p$ is the mass source. The volumetric mass source term for liquid phase is

$$Q_L = R_{PT,w,L} MW_w \tag{42}$$

where $R_{PT,w,L}$ is the source term for water in liquid phase,

$$R_{PT,w,L} = -\frac{a_s k_{MT,L}}{RT}\left(P_L - P_{L,M}\right) + k'_{MT}(RH - 100\%)\left[\mathrm{H}_0\left(\frac{P_L}{P^{ref}}\right) + \mathrm{H}_0(RH - 100\%)\right] \tag{43}$$

where the first term represents mass transfer between liquid water and the ionomer in the catalyst layers, and the second term defines water evaporation/condensation in catalyst layers and porous layers. The mass-transport coefficient  is an arbitrarily larger number, $k'_{MT} = 10^7 mol\, m^{-3} s^{-1}$ and $\mathrm{H}_0(x)$ is a Heaviside step function employed to ensure relative humidity does not exceed 100% when liquid water is present $(p_L > 0)$.

Gas diffusion in the CLs and porous layers is described by Stefan-Maxwell equations:

$$N_j = -\rho_g D_j^{\mathrm{eff}} \nabla\omega_j - \rho_G \omega_j D_j^{\mathrm{eff}} \frac{\nabla M_n}{M_n} + \rho_j u_G \tag{44}$$

where $\rho_g$ is the gas mixture density, $\omega_j$ is the mass fraction species $j$, $u_G$ is the average gas velocity calculated using Darcy's equation (Eq. 37). $M_n = \left(\sum_j \frac{\omega_j}{M_j}\right)^{-1}$ is the average molecular weight of the mixture. Additionally, the mole fractions of all species, $x_j$ sum to one.

$$\sum_j x_j = 1 \tag{45}$$

where $x_j = \frac{\omega_j M_n}{M_j}$. The effective diffusivity is defined as

$$D_j^{\mathrm{eff}} = \varepsilon_G^{1.5}\left(\frac{1}{D_j^m} + \frac{1}{D_j^k}\right)^{-1} \tag{46}$$

with the mixture-averaged diffusivities given by

$$D_j^{\mathrm{m}} = \frac{1 - \omega_\mathrm{j}}{\sum_{q \neq j} \frac{x_q}{D_{jq}}} \tag{47}$$

where $D_{jq}$ is the binary gas-phase diffusivity estimated following derivation by Fuller *et al.*[44].

$$D_{j,q} = \frac{10^{-3}T[K]^{1.75}(M_i[\text{g mol}^{-1}]^{-1} + M_i[\text{g mol}^{-1}]^{-1})^{0.5}}{p_G[\text{atm}]\left(v_{p,j}^{0.33} + v_{p,q}^{0.33}\right)^2} \tag{48}$$

where $v_{p,j}$ is the diffusion volume of species $j$. The Knudsen diffusivity,

$$D_j^K = \frac{2r_{\text{pore,m}}}{3}\sqrt{\frac{8RT}{\pi M_i}} \tag{49}$$

relates to the transport in small pores, where $r_{\text{pore,m}}$ is the averaged pore radius for medium $m$. The volumetric mass source term for gas phase includes water vapor, $CO_2$, and gaseous $CO_2$R products:

$$Q_G = R_{PT,w,G}MW_w - R_{PT,CO_2,I}MW_{CO_2} + \sum_{k=O_2,H_2,CO,C_2H_4} R_k MW_k \tag{50}$$

where $R_{PT,w,G}$ is the source term for water in gas phase,

$$R_{PT,w,G} = -a_s k_{MT,V}\left(\frac{RH}{100} - a_w\right) - k'_{MT}(RH - 100\%)\left[\mathrm{H}_0\left(\frac{P_L}{Pref}\right) + \mathrm{H}_0(RH - 100\%)\right] \tag{51}$$

where the first term represents mass transfer between vapor water and the ionomer in the catalyst layer, and the second term defines water evaporation/condensation in the porous layers.

The rate of $CO_2$ mass transfer into the ionomer is expressed as

$$R_{PT,CO_2,I} = a_s k_{MT,j}\left(c_{CO_2}^{eq} - c_{CO_2}\right) \tag{52}$$

where $k_{MT,j}$ is the mass-transfer coefficient for gaseous $CO_2$ entering the ionomer phase, defined as

$$k_{MT,j} = \frac{D_{CO_2,w}}{\delta_{TF}} \tag{53}$$

where $D_{CO_2,w}$ is the intrinsic $CO_2$ diffusivity in the water, $\delta_{TF}$ is the thickness of the electrolyte film covering the pore walls, $c_{CO_2}^{eq}$ is the concentration of $CO_2$ in equilibrium with its concentration external to the ionomer:

$$c_{CO_2}^{eq} = p_{CO_2} H_{CO_2} \tag{54}$$

where $p_{CO_2}$ is the partial pressure of $CO_2$, $H_{CO_2}$ is Henry's constant for $CO_2$ solubility. The mass transport of $CO_2$ within the ionomer/liquid phase is described using the diffusion equation,

$$\mathbf{N}_{CO_2} = -D_{CO_2,w} \nabla c_{CO_2} \tag{55}$$

and the conservation of $CO_2$ within the membrane/ionomer can be expressed as

$$\nabla \cdot \mathbf{N}_{CO_2} = R_{PT,CO_2,I} \tag{56}$$

The intrinsic properties of $CO_2$ in water are listed in Table S3 in the SI. Finally, the $CO_2$ utilization efficiency defined as

$$1 - \frac{\int_{aCL} R_{PT,CO_2,I} \cdot MW_{CO_2} \;\; dx}{\int_{cCL} R_{PT,CO_2,I} \cdot MW_{CO_2} \;\; dx} \tag{57}$$

where $MW_{CO_2}$ is the molecular weight of $CO_2$. While the denominator represents the electrochemically consumed $CO_2$ at the cathode, numerator represents the released and crossover $CO_2$ at the anode.

## 2.6 Conservation of Energy and Heat transfer

The temperature profile in the system calculated by solving the conservation of energy and heat transfer equation,

$$\rho^{eff} c_p^{eff} \frac{\partial T}{\partial t} + \rho_f c_{p,f} u \cdot \nabla T - k_T^{eff} \nabla T = Q_{CT} + Q_B + Q_{PT} + Q_j \tag{58}$$

where $\rho^{eff}$ is effective density, $\rho_f$ is the density of the fluid, $c_p^{eff}$ is effective specific heat capacity, $c_{p,f}$ is the heat capacity of the fluid, $u$ represents the fluid velocity, calculated using Darcy's equation (Eq. 37), $k_T^{eff}$ is the effective thermal conductivity. The effective properties are calculated based on the volume fractions and properties of the solid and fluid phases:

$$k_{T,m}^{eff} = \varepsilon_s k_{T,s} + \varepsilon_f k_{T,f} \tag{59}$$

The source terms $Q_{CT}$, $Q_B$, $Q_{PT}$, $Q_j$ correspond to charge-transfer reactions, homogeneous bulk reactions, phase transfer, and Joule heating, respectively. Heat generation from electrochemical reactions, including both reversible and irreversible terms, is given by:

$$Q_{CT} = \sum_j (i_j \eta_j + i_j \Pi_j) \tag{60}$$

where $\Pi_j$ is the Peltier coefficient for reaction $j$, listed in Table S3 in the SI. Heat generation due to the homogenous reactions is given by

$$Q_B = \sum_n \Delta H_n \left( k_n \prod_{v_{j,n} < 0} c_j{}^{-v_{j,n}} - \frac{k_n}{K_n} \prod_{v_{j,n} > 0} c_j{}^{v_{j,n}} \right) \tag{61}$$

where $\Delta H_n$ is the enthalpy of the reaction $n$ provided in Table S3 in the SI. Heat generation via phase transfer is given by

$$Q_{PT} = -k'_{MT}(RH - 100\%) \left[ H_0 \left( \frac{P_L}{Pref} \right) + H_0(RH - 100\%) \right] \Delta H_{vap} \tag{62}$$

where $\Delta H_{vap}$ is the enthalpy of the water. Joule heating resulting from electric and ionic currents,

$$Q_J = \frac{i_s^2}{\sigma_s^{eff}} + \frac{i_l^2}{\sigma_l^{eff}} \tag{63}$$

All intrinsic thermal properties are provided in Table S3 in the SI.

## 2.7 $C_2H_4$ Production Cost Calculation

The molar production rate of $C_2H_4$ is

$$\dot{n}_{C_2H_4} = \frac{i_{C_2H_4}}{nF} \tag{64}$$

where $i_{C_2H_4}$ is partial current density of $C_2H_4$, $n$ is the number of electrons consumed, and $F$ is the Faraday's constant. Electrical power consumed to produce $C_2H_4$ is

$$P = i_{total} * V_{cell} \tag{65}$$

where $i_{total}$ is the total current density, $V_{cell}$ is the cell voltage. Thus, the energy consumption for kg of $C_2H_4$ can be calculated as:

$$\text{The electric energy cost for kg of } C_2H_4 = \frac{P}{\dot{n}_{C_2H_4} * MW_{C_2H_4}} * 0.01\,\$\,kWh^{-1} \tag{66}$$

where $MW_{C_2H_4}$ is the molecular weight of $C_2H_4$. Assuming the electric price 0.01 \$ $kWh^{-1}$, Electric energy cost for 1 kg of $C_2H_4$ (\$ $kg^{-1}$) can be found.

## 2.8 Boundary Conditions

The boundary conditions for the above governing equations are adopted from our previous work and listed in Table 1. The solid-phase potential (electrical potential) $\phi_s$ is set to the operation cell potential at the anode channel (aCH) and aDL boundary while the cathode channel (cCH) and cDL boundary is arbitrarily grounded. The gas pressure $p_G$ is set to the atmospheric pressure (101 atm) at both aCH|aDL and cCH|cDL boundaries. While the liquid pressure $p_L$ is set to 1 atm at the aCH|aDL boundary (liquid feed), it is set to 0 atm at the cCH|cDL boundary assuming no liquid present at the cathode. The mass fractions are assumed to be fixed at the aCH|aDL and cCH|cDL boundaries with the cathode being 100% RH $CO_2$ and anode is 100% RH $N_2$. No-flux boundary condition are implemented at the aDL|aCL and cGDL|cCL boundaries both for both electroactive species and water. The operating temperature, $T_0$ is set at both aCH|aDL and cCN|cDL boundaries.

**Table 1** Boundary conditions implemented in the model.

| | aCN\|aDL | aCN\|cDL | aDL\|aCL | aDL\|aCL | Governing equation |
|---|---|---|---|---|---|
| $\phi_s$ | $V_{cell}$ | $0\,V\ (ground)$ | | | 22 |
| $p_G$ | $1\ atm$ | $1\ atm$ | | | 37 |
| $p_L$ | $1\ atm$ | $0\ atm(No\ liquid)$ | | | 37 |
| $\omega_i$ | $\omega_w = \frac{P_w^{vap} M_w}{p_G M_A}$<br>$\omega_{N_2} = 1 - \omega_w$<br>$\omega_{i \neq w, N_2} = 0$ | $\omega_w = \frac{P_w^{vap} M_w}{p_G M_A}$<br>$\omega_{CO_2} = 1 - \omega_w$<br>$\omega_{i \neq w, CO_2} = 0$ | | | 44 |
| $T_{temp}$ | $T_0$ | $T_0$ | | | 58 |
| $c_i$ | | | $\nabla \cdot \mathrm{N}_i = 0$ | $\nabla \cdot \mathrm{N}_i = 0$ | 24 |
| $\mu_0$ | | | $\nabla \cdot \mathrm{N}_w = 0$ | $\nabla \cdot \mathrm{N}_w = 0$ | 32 |

## 2.9 Computational Methods

The model was developed in COMSOL Multiphysics® 6.0 software. The modelling domain featured a maximum element size of 0.1 μm. Due to the complexity of the physics in the aCL, AEM, and cCL, the element size was reduced to 0.01 $\mu m$ within these domains. Additionally, the element size was further refined to $1 \times 10^{-4}$ μm at the aCL|AEM and AEM |cCL boundaries to capture the sharp concentration gradients. The governing equations were solved sequentially using multiple study steps. Initially, Darcy's equation was solved for both liquid and gas phases (Eq. 37) to determine pressure and velocity. This was followed by solving the Stefan-Maxwell equations (Eq. 44) for the mass transport of gases. Subsequently, an initialization study was performed for charge transport equations (Eq. 9, Eq. 22 and 24) and water transport equations (Eq. 32), which were solved sequentially in individual study steps. The conservation of energy and heat transfer equations (Eq. 58) were solved next to determine the temperature field. Finally, the entire model, including all governing equations and electrochemical heating multiphysics coupling, was solved in the last step. The auxiliary sweep feature was employed to ramp the cell potential from 1.9 V to 4 V with a 1 mV potential step. Based on the physics involved, the default COMSOL solvers,

PARDISO and MUMPS, were used throughout the simulation with a relative tolerance of $1 \times 10^{-4}$.

## 2.10 Experimental Methods

### 2.9.1 Catalyst Electrode Preparation

The catalyst-layer ink was fabricated by bath sonicating (CPX2800H, Branson) 50 mg Cu nanoparticles (NPs) (Sigma-Aldrich, average size of 25 nm) and 50 mg of Nafion solution (5 wt%, Ion power, D521) in 25 mL of isopropyl alcohol for 1 hour. The Cu NP catalyst electrode was fabricated using a Sono-Tek spray coating of as-prepared catalyst ink on the porous carbon gas-diffusion layer (Sigracet 39BB) at 120 kHz sonication with 0.2 $cm^{-3}$ $min^{-1}$ of ink injection rate. The catalyst loading amount was confirmed to be 0.5 mg $cm^{-2}$ as measured by X-ray fluorescence spectroscopy (XRF, Bruker).

The anode catalyst ink was prepared by 1 h bath sonicating a mixture of 150 mg of $IrO_2$ powder (Alfa Aesar) and 0.6 g of PiperION A ionomer solution (5 wt%, Versogen) in 1 mL deionized water and 9 mL $n$-propanol. The anode was fabricated by hand spray of as-prepared $IrO_2$ catalyst ink on a laser-cut 5 $cm^2$ platinized Ti porous-transport layer (Mott Corp.) using an airbrush gun (RichPen GP-1). $IrO_2$ loading amount was controlled to be 1.3 mg $cm^{-2}$, which was measured by an electronic balance.

### 2.10.2 Electrochemical $CO_2$ Reduction

A $5\,cm^2$ single-cell MEA electrolyzer hardware (Fuel Cell Technology, FCT) with a single serpentine-channel graphite flow field on the cathode and a single-serpentine channel Pt-coated Ti flow field on the anode was used for the $CO_2$RR test. The cell was assembled by stacking as-prepared anode, AEM (PiperION-A, 40 μm thickness, Versogen) and cathode and subsequently compressed to 40 in-lbs in 10 in-lbs increments to form the MEA. The AEM was pre-activated in 1 M KOH solution for at least 24 h and was rinsed by deionized water before assembly. 20% and 0% compression of the cathode GDL and anode PTL, respectively, was achieved by controlling the ethylene tetrafluoroethylene (ETFE) gasket thickness.

During the $CO_2R$ test, humidified $CO_2$ gas (99.999%, Airgas) at ambient temperature continuously flowed in the cathode flow field at 200 $cm^3$ $min^{-1,}$ and 0.5 M $KHCO_3$ electrolyte was circulated with a rate of 20 $cm^3$ $min^{-1}$ using a peristaltic pump. The effluent gas was connected to gas chromatograph (SRI Company) through a water trap containing 20 mL of deionized water. Anolyte volume was confined to 20 mL $CO_2R$ was performed by applying step-wised cell potential from 3 to 4 V in 0.5 V increments. Gas-chromatography data were collected after 5 min and 20 min of applying each potential. After 30 min at each potential, 0.2 mL of catholyte and anolyte were collected and mixed with 0.1 mL of $D_2O$ (99.9 atom% D, Sigma Aldrich), containing 10 mM of dimethyl sulfoxide (DMSO) as an internal standard for quantification of liquid products by proton nuclear magnetic resonance ($^1$H-NMR, Bruker Ascend 500 MHz instrument). The FE of each product was calculated using

$$FE_i = \frac{n_i F j_i}{i_{tot}} \times 100 \tag{67}$$

where $i_{tot}$ is the total current, $j_i$ is the moles of product $i$ generated per second quantified by gas chromatography or $^1$H NMR, $F$ is Faraday's constant, and $n$ is the number of electrons to generate product $i$ via $CO_2RR$.

# 3. Results and Discussion

Figure 1a shows an expanded view of the MEA cell used for the acquisition of experimental data and an illustration of the through-plane 1-D computational domain. The cell consists of a flow channel through which $CO_2$ saturated with water vapor is supplied to the cathode. A similar flow channel is used to supply liquid electrolyte (0.5 M $KHCO_3$) to the anode. The cathode diffusion layer (cDL) and the anode diffusion layer (aDL) provide pathways for reactants/products to move from/to the flow channels to/from the catalyst layers (CL). The cDL is coated with a layer of Cu nanoparticles (5 μm thick, 0.6 mg $cm^{-2}$) and the aDL is coated with a layer of $IrO_2$ nanoparticles (5 μm thick, 1.3 mg $cm^{-2}$). An anion-conducting membrane (Versogen PiperION, 40 μm thick) was placed between the aDL and cDL such that the CL of each aDL and cDL were in direct contact with the membrane. Figure 1b illustrates the regions of the cell that were simulated using a 1-D model of transport and reaction occurring in the aDL, anode CL, membrane, cathode CL and cDL.

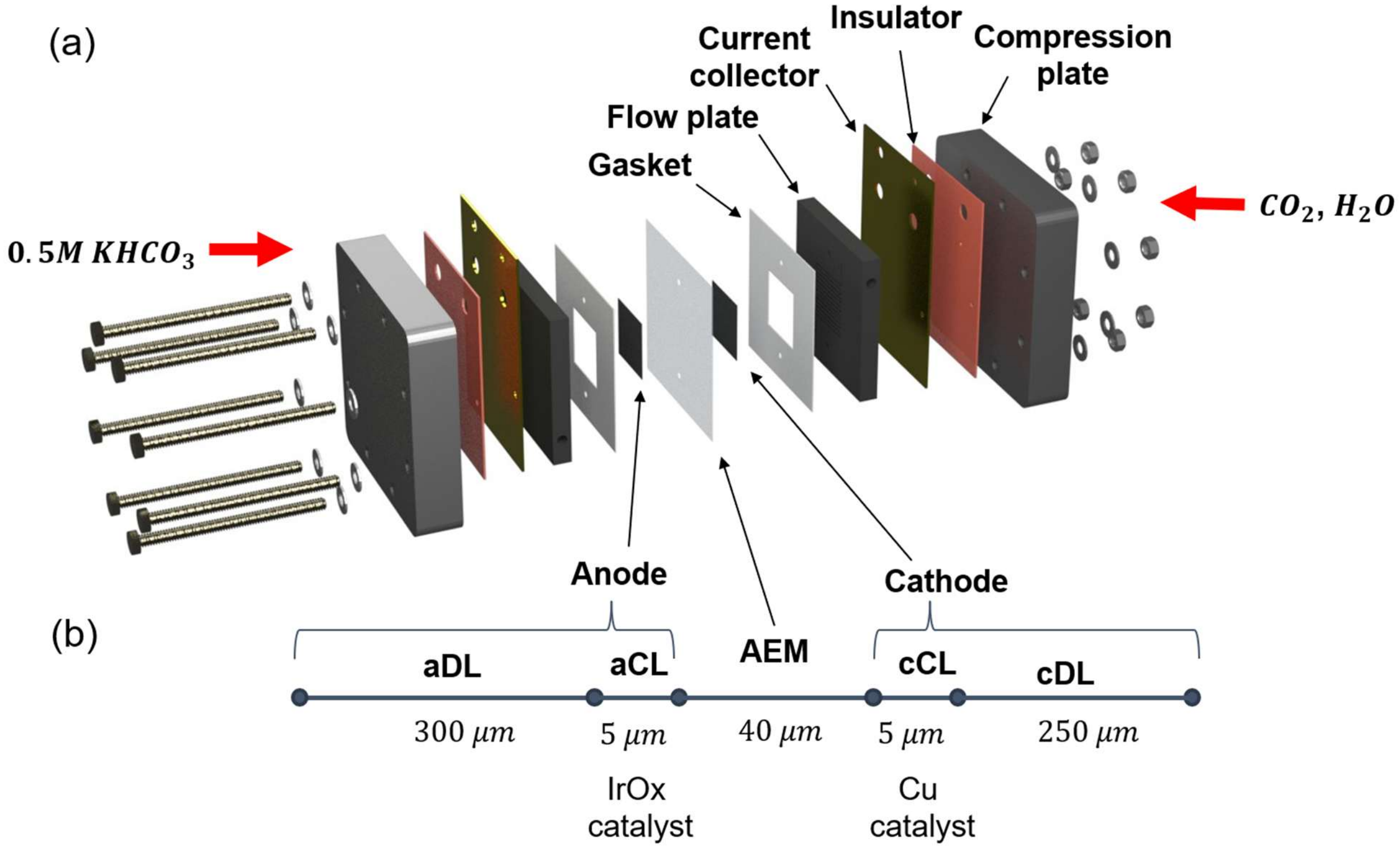

**Figure 1.** (a) The exploded view of 5 $cm^2$ $CO_2$ reduction membrane-electrode-assembly cell and (b) 1-D computational domain.

Figure 2a and 2b show the experimentally measured total current density and FEs for $H_2$, CO, $C_2H_4$, HCOOH, $C_2H_5OH$, and $C_3H_7OH$ versus cell potential. These data were obtained by analyzing the composition of the gas products analysis from cathode outlet and quantifying the liquid products in both circulating anolyte and samples in cathode water trap. Since $C_2H_5OH$ and $C_3H_7OH$ diffuse to the anode side of the MEA, where they are probably oxidized to $CH_3COO^-$ and $CH_3CH_2COO^-$, respectively, the rates at which these anions are formed together with the rates of the corresponding alcohols formed in the cathode CLs are lumped. We also note that some of the $HCOO^-$ anions formed at the cathode crossover through the AEM to the anode chamber. Since these species were formed in the cathode, the rate of the appearance of these anions are also lumped into the rate at which they are formed in the cathode. The experimental data shown in Figure 2a and 2b represent the average value over three identical MEA assemblies for which the compositions of all components were the same; the variation in each quantity measured is indicated by the error bars.

Best-fit predictions of the data obtained from our 1-D model are shown in Figure 2c and 2d. To do so, we varied the exchange current density and charge-transfer coefficient for formation of each product, while holding all other parameters constant at values obtained from the literature. Details of the fitting procedure and a list of parameters taken from the literature are given in the Supporting Information. The simulated total current densities agree with the three experimental data reasonably well, using only exchange-current density and charge-transfer coefficients for each product as fitting parameters. The observed levels of agreement are probably a consequence of using Tafel expressions for the reaction kinetics based on experiments conducted with planar copper electrodes and not nanoparticles. We note that recent work conducted simulating the formation of CO in MEA containing Ag particles has shown that the kinetics are more appropriately described by Marcus-Hush-Chidsey (MHC) kinetics [31]; however, similar studies have not been reported for $CO_2R$ to CO on Cu. Additionally, salt precipitation at the cathode side due to cation transport and accumulation (e.g., $K^+$) can affect local reaction kinetics and transport properties under prolonged operation. However, the scope of this study was limited to evaluating how $CO_2R$ performance varies with operating conditions, and long-term stability effects are beyond the scope herein. It should also be noted that the model incorporates energy conservation, resulting in a minimal temperature gradient of only 3 to 4℃ within the MEA. The center of the MEA, which contains AEM is hotter than the boundaries, where the temperature is equal to the ambient temperature.

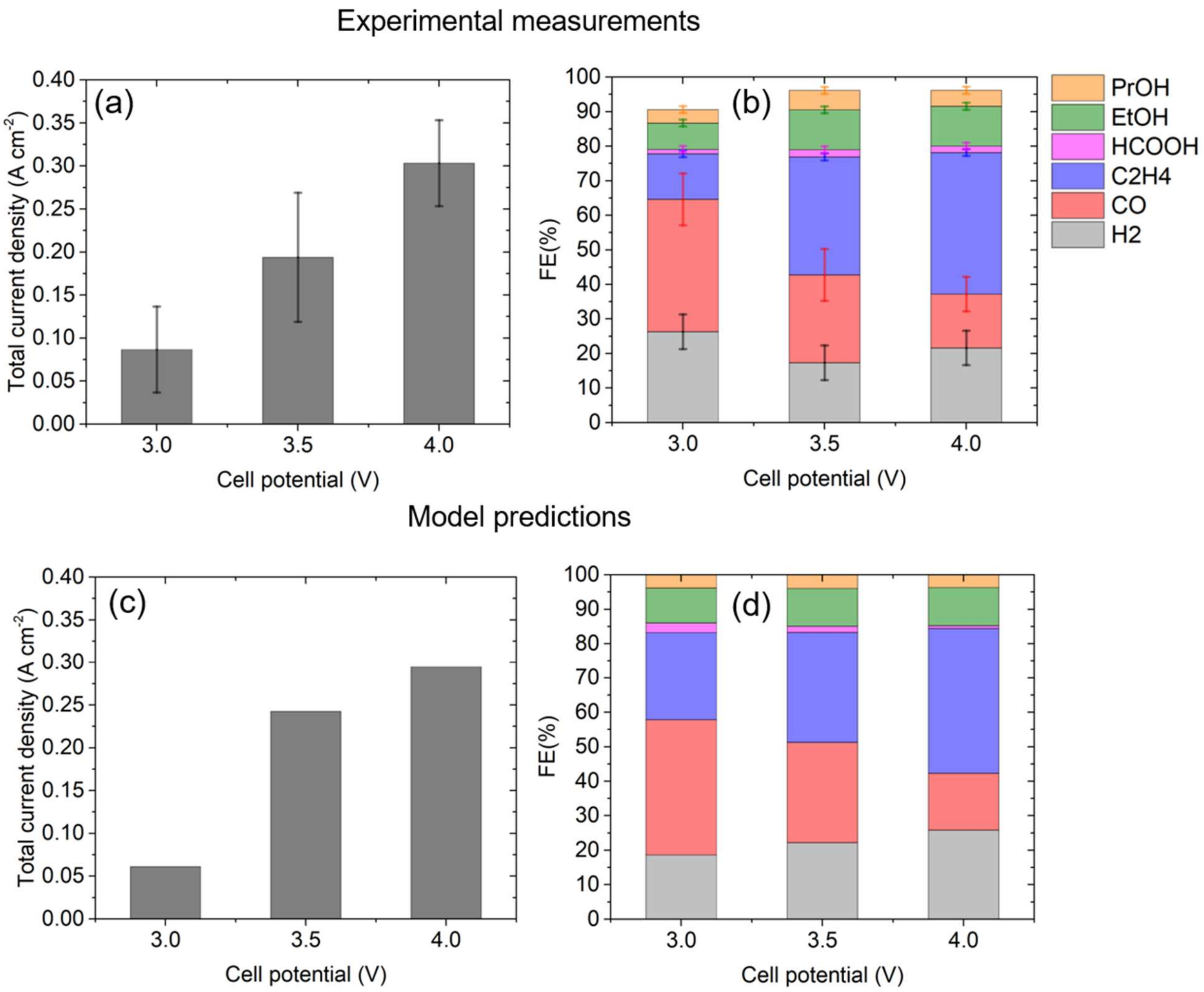

**Figure 2.** Experimentally measured and computationally predicted current densities and FEs for $H_2$, CO, $C_2H_4$, HCOOH, $C_2H_5OH$, $C_3H_7OH$ at 3 V, 3.5 V, and 4 V cell voltage.

The MEA model for the base case is used to obtain the overall polarization curve (Figure 3a) by varying the cell voltage between 1.9 V and 4 V in 1 mV potential step. The model is then used to obtain an applied voltage breakdown (see SI for details) for current densities of 0.05, 0.15, and 0.25 A $cm^{-2}$. Figure 3b shows the total cell voltage divided into five components: the anode equilibrium potential, the cathode equilibrium potential, the anode kinetic overpotential, the cathode kinetic overpotential, and the ohmic potential loss. The equilibrium potential represents the minimal potential needed to drive half-cell reactions occurring at the anode and cathode. The equilibrium potential changes relative to the standard thermodynamic potential value in the standard state (298 K, 1 M concentration of ionic species, and a water activity of 1.0) because of differences in the reaction conditions (temperature, pressure, and pH) at each electrode relative to those for the standard state. The equilibrium potentials for the anode and cathode half-cell reactions were calculated considering the potential change due to changes deviations of the conditions at the anode and cathode surface from those in the standard state (*i.e.*, Nernstian corrections). The kinetic overpotential is the potential required to drive charge-transfer reactions to achieve a desired current density. The ohmic overpotential is associated with the potential needed to conduct ions through the membrane and the ionomer in each CL. Ohmic overpotential also consist of an overpotential due to the electron conduction in the solid phase (aDL, cDL, and CLs); however, this overpotential loss is minor, only 5 mV. Figure 3b shows that with increasing current density the cell potential increases mainly due to increases in the anode and cathode overpotential and the ohmic overpotential. Minor changes also occurred in the cathode equilibrium potential due to changes in both pH and product distribution changes at higher current densities. The voltage-breakdown for all current densities is reported in Figure S2.

Simulated product distributions for three different current densities are shown in Figure 3c. While the FEs for $H_2$ and $C_2H_4$ increase with increasing current density, the FE for CO decreases. The FE for HCOOH becomes negligible at higher current density, whereas the FEs for $C_2H_5OH$ and $C_3H_7OH$ level out and together account for > 40% of the total FE. Simulated product distributions at any current density or cell voltage can be found in Figure S3a and S3b in the SI, respectively. To understand better the factors affecting $C_2H_4$ formation, we examined the spatial variation in the FE for $C_2H_4$, the electrolyte potential ($\phi_l$), the electrode reaction rate, and the pH within the cathode CL, which are shown in Figure 3d, 3e, 3f and 3g, respectively. A strong correlation is observed between the FE for $C_2H_4$ and the electrolyte potential, with the FE for $C_2H_4$ increasing with electrolyte potential. The electrolyte potential and the $C_2H_4$ FE is highest at the AEM|cCL interface and decreases as one moves further away from the AEM. Similarly, Figure 3f shows that the most active portion of the cathode CL is that closest to the AEM, and the activity decreases rapidly (especially at 0.25 A $cm^{-2}$) as one moves away from the AEM|cCL. These results suggest that $OH^-$ ionic transport is the most limiting factor for the reactions. The $OH^-$ activity also plays a role in product selectivity due the effect of pH on the Tafel kinetics for the hydrogen-evolution reaction (HER). Higher pH suppresses HER but has no impact on the rate of $CO_2R$ to $C_2H_4$ and, hence, higher pH increases the FE for $C_2H_4$. Figure 3g shows that pH and its gradient increases at higher current densities. The local pH is governed by a complex interplay between the generation, consumption, and transport of $OH^-$ ions. At higher current densities, we observed a decrease in water activity due to increased water consumption (see Figure S22 in SI), which in turn limited $OH^-$ transport. This limitation led to more pronounced shifts in the local pH. It is notable, though, that the impact of pH on product selectivity is relatively small compared to the effect of the potential. A higher potential promotes the production of both $H_2$ and $C_2H_4$, driven by their large charge-transfer coefficients; for further details, refer to the Supporting Information (SI).

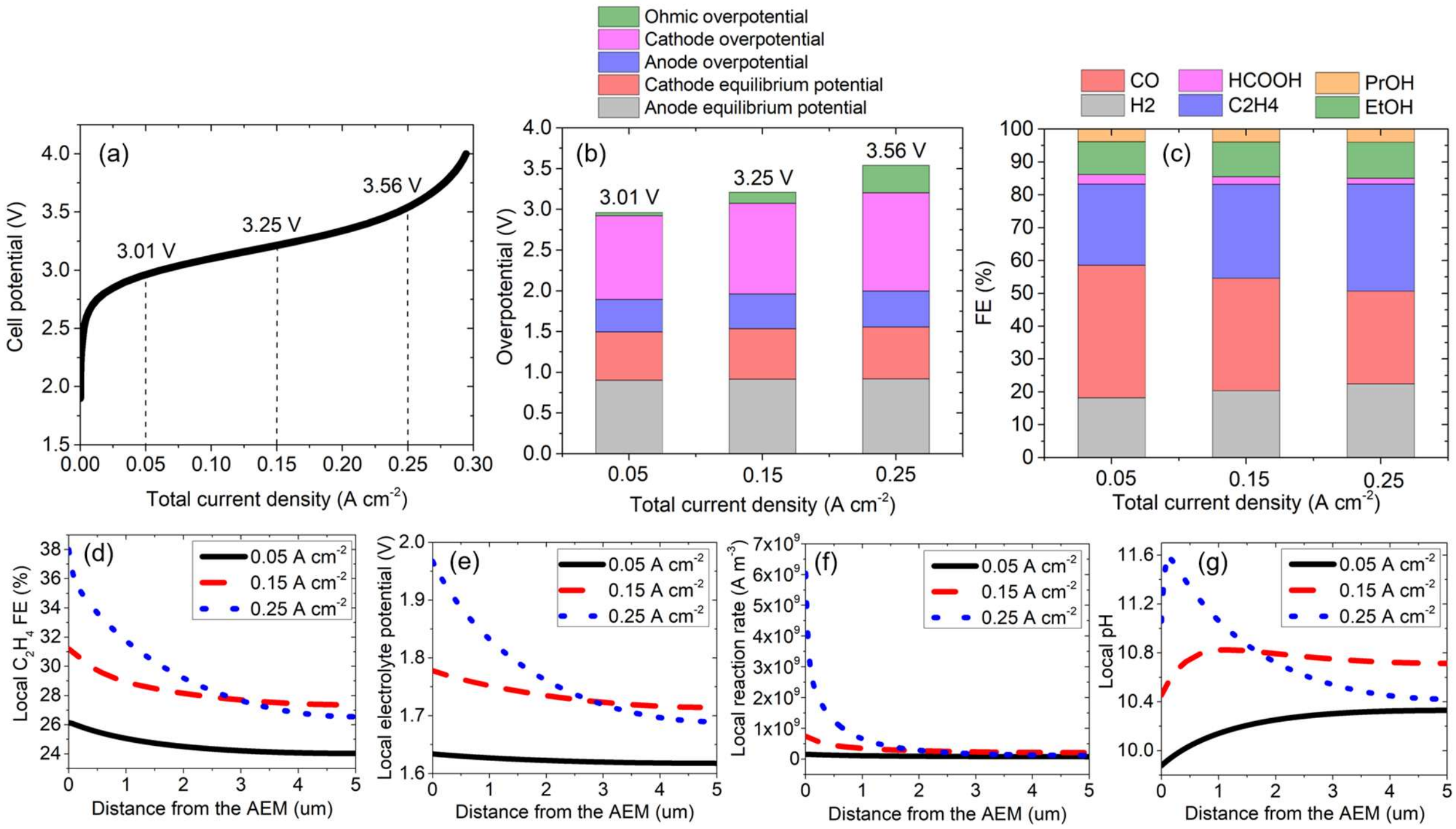

**Figure 3.** The model predictions for the base case (a) overall polarization curve (b) applied-voltage breakdown (c) product distribution taken from model analysis (d) local $C_2H4$ FE distribution (e) local electrolyte potential distribution (f) local electrode reaction source distribution (g) local pH distribution.

Next, we used the model to explore the effects of anode and cathode CL thickness and ECSA, membrane thickness and conductivity, water transport, and $CO_2$ concentration on the formation of $C_2H_4$. We found that the cell potential required for a given total current density and the FE to $C_2H_4$ are affected most significantly by the cathode CL thickness, ECSA, and membrane thickness (see Figure S4 and S5 in the SI). The influence of each of these variables is described below, and the effects of all other variables are presented in the Supporting Information.

Figure 4a illustrates the polarization curves for cathode CL thicknesses of 2.5, 5, and 10 μm. As the thickness increases, the potential for current densities between 0 and 0.2 A $cm^{-2}$ decreases but has relatively little effect at higher current densities. Figure 4b compares individual overpotential contributions at 0.15 A $cm^{-2}$ current density for all CL thicknesses. The potential decreases with increasing cathode CL thicknesses. This is a consequence of two main factors: (i) a decrease in the cathode potential and (ii) an increase in the ohmic potential loss. There is also a minor decrease observed on the cathode equilibrium potential due to the lower pH for thicker cathode CL. A more detailed voltage-breakdown comparison can be found in Figure S6 in the SI. Figure 4g shows that the local pH is lower for thicker cathode CLs, resulting in a smaller Nernstian shift in the cathode equilibrium potential. This decrease in pH is primarily due to intensified buffering reactions. In thicker CLs, longer diffusion pathways hinder $OH^-$ transport, leading to local accumulation. The elevated $OH^-$ concentration promotes buffering reactions with $CO_2$, forming $HCO_3^-$ and $CO_3^{2-}$. As a result, these enhanced buffering processes reduce the final local pH in thicker catalyst layers. These findings are consistent with previously reported trends in the literature [45,46]. Figure 4c illustrates the effect of cathode CL thicknesses on the FEs for all products at 0.15 A $cm^{-2}$. For all other current densities, refer to Figure S7 in the SI. The model predicts a slightly higher FE for $C_2H_4$ for a thinner cathode CL, which is mainly due to local potential gradients at the cathode CL. As mentioned above, the potential has a significant impact on the product FEs, and as Figure 4e shows, a thinner cathode CL has a relatively higher electrolyte potential, and thus higher $C_2H_4$ production rates as seen in Figure 4d. Figure 4f shows that the thinner cathode CLs are more active than the thicker CLs at the same current density (*i.e.* fluxes) due to the relatively more alkaline environment (see Figure 4g), which facilitates $C_2H_4$ production.

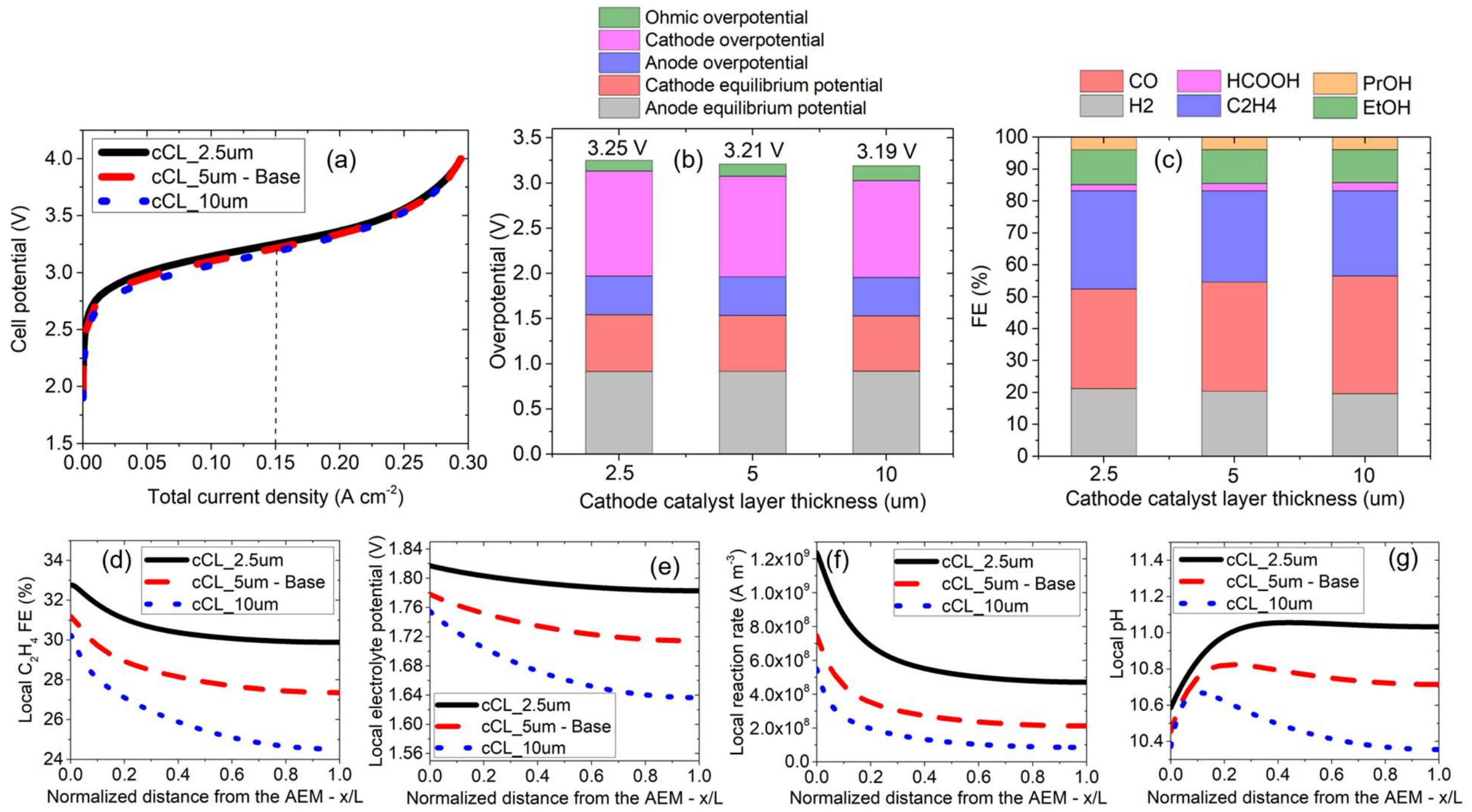

**Figure 4.** Predicted (a) overall polarization curve (b) applied-voltage breakdown (c) product distribution taken from model analysis (d) local $C_2H4$ FE distribution (e) local electrolyte potential distribution (f) local reaction rate distribution and (g) local pH distribution for different cathode CL thicknesses.

The cathode ECSA was found to have the largest impact on cell potential at a given current density and on the FE for $C_2H_4$ formation. We note that for fixed Cu weight loading, the ECSA can be changed by increasing or decreasing the Cu dispersion and by changing the ionomer-to-catalyst ratio while keeping all other variables the same [12]. Polarization curves are compared in Figure 5a for values of the ECSA that are smaller or larger by a factor of 10 than that for the base case. Increasing the cathode ECSA notably decreases the cell potential for a current density of 0.15 A $cm^{-2}$. Figure 5b shows that this change is due to the combined effects of changes in the cathode equilibrium potential and the cathode overpotential. The cathode equilibrium potential increases slightly with increasing ECSA due to the increase in CO production as seen in Figure 5c. The change in cathode equilibrium potential is offset by the decrease in cathode overpotential. It is notable that increasing the cathode ECSA has a small effect on both the anode overpotential and the ohmic resistance of the cell. Figure 5c compares the total FE for all products as a function of cathode ECSA. It is observed that the formation of $C_2H_4$ is greater for the lower cathode ECSA. Figure 5e shows that the local potential distribution is noticeably higher within the cathode CL for the lower ECSA. Figure 5d shows that the local FE for $C_2H_4$ exhibits a very similar trend to the local electrolyte potential distributions in the cathode CL. The pH impact on the FE of $C_2H_4$ is the same for all ECSAs due to the similar reaction and pH distributions (Figure 5f and 5g) at a current density of 0.15 A $cm^{-2}$. For more information on both individual overpotential contributions and product distributions as function of cathode ECSA, refer to the Supplementary information Figure S8 and Figure S9.

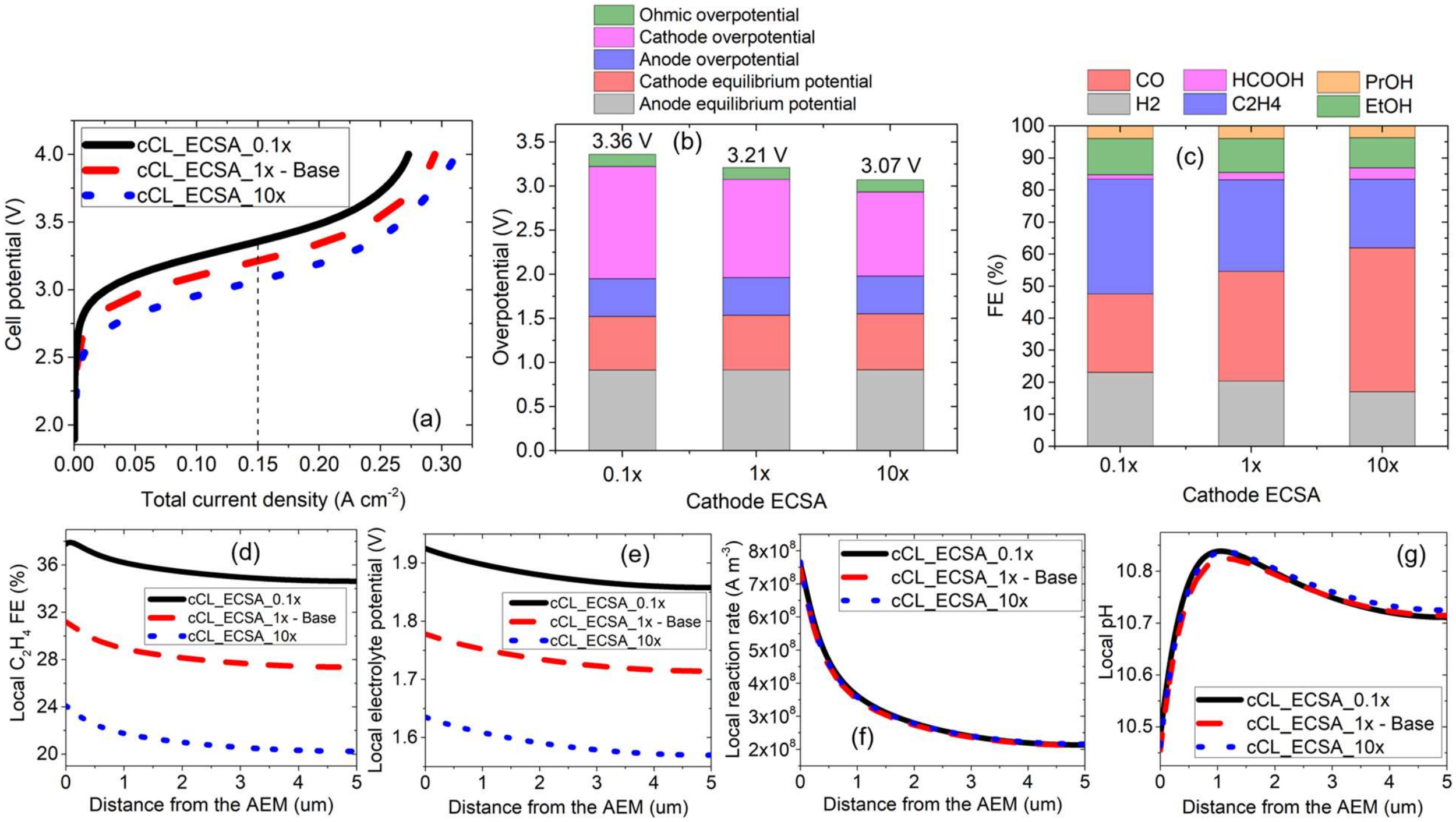

**Figure 5.** Model predictions as a function of different cathode specific ECSA: (a) overall polarization curve (b) applied-voltage breakdown (c) product distribution taken from model analysis (d) local C2H4 FE distribution (e) local electrolyte potential distribution (f) local electrode reaction source distribution and (g) local pH distribution.

We also explored the consequences of changing the AEM thickness by keeping all other MEA properties the same as in the base case. Figure 6a shows the polarization curves for three cases: the baseline case for which the AEM thickness is 40 μm and for AEM thicknesses of 20 and 60 μm. As expected, the cell potential decreases for a given total current density as the AEM thickness decreases. This is mainly due to the decrease in ohmic overpotential with decreasing AEM thickness. The effects of AEM thickness on the anode and cathode equilibrium potentials and overpotentials are relatively small at a current density of 0.15 A $cm^{-2}$ as seen in Figure 6b. The effects of AEM thickness on the FEs for all products are given in Figure 6c. With increasing AEM thickness, the FE for $C_2H_4$ increases slightly due to the increase in the local potential at the cathode CL (see Figure 6e). Similar trends have been reported in the literature, supporting the accuracy of the model [47]. However, the thicker AEM becomes ion-transport limited due to the higher ohmic overpotential for transporting ions across the thicker membrane, which is conflated with the lower water content. This limitation causes a nonuniform reaction distribution at the cathode CL. Figure 6f shows that the active portion of the CL shifts towards AEM for a cell with thicker AEM due to the exacerbated ion transport limitations. As a result of this, a relatively higher pH is observed (see Figure 6g). For more information on both individual overpotential contributions and product distributions as function of AEM thickness, see Figure S10 and Figure S11 in the SI.

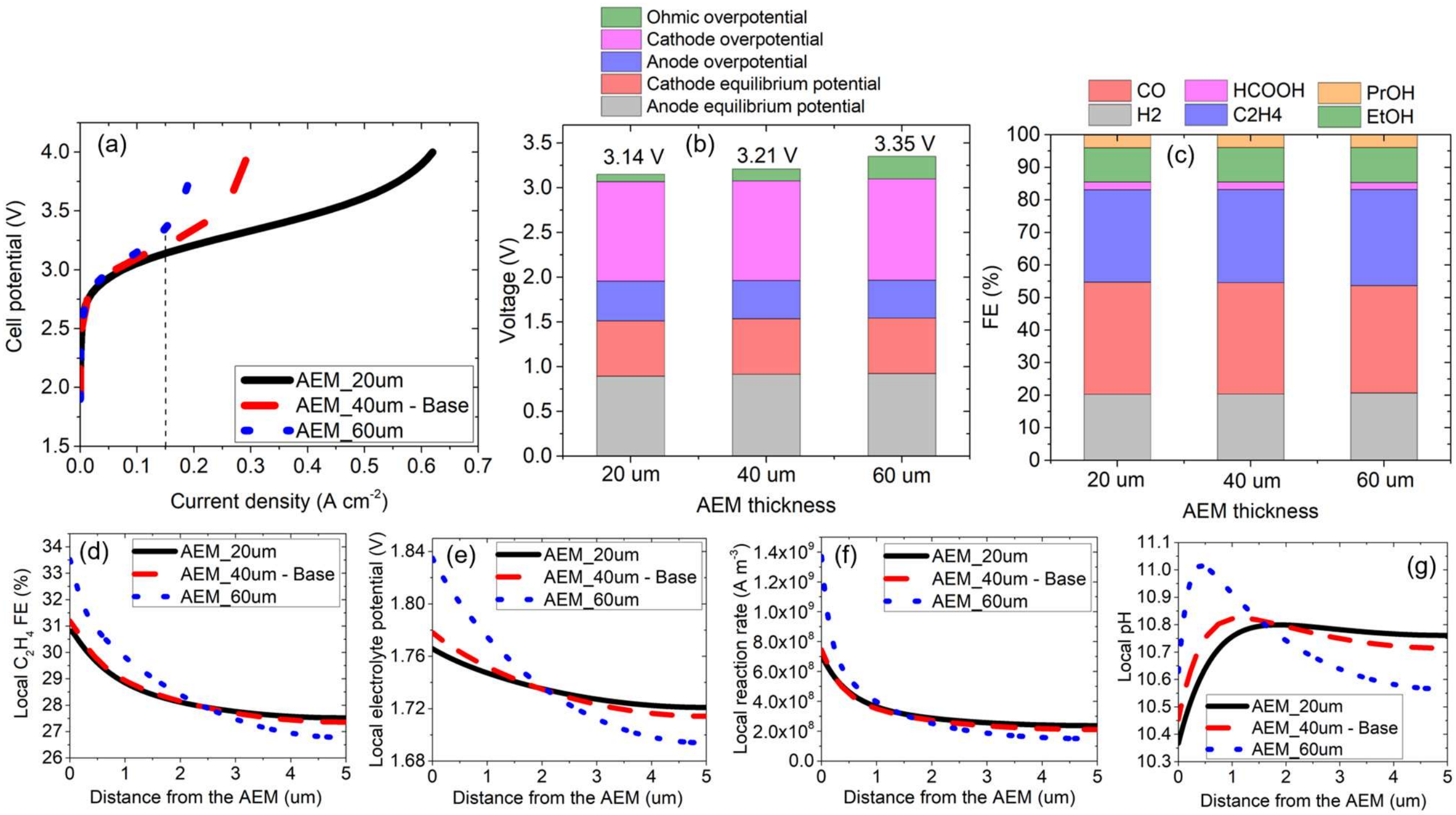

**Figure 6.** Model predictions for different AEM thicknesses: (a) overall polarization curve (b) applied-voltage breakdown (c) product distribution taken from model analysis (d) local C2H4 FE distribution (e) local electrolyte potential distribution (f) local electrode reaction source distribution and (g) local pH distribution.

The effects of anode CL thickness, anode ECSA, AEM conductivity, and $CO_2$ partial pressure, on the polarization curve and the FEs for all products are reported in the SI. Increasing AEM conductivity and water transport in the AEM and ionomer portion of the CLs decreases cell potential due to decreasing the ohmic overpotential; however, the impact on the FE for $C_2H_4$ is negligibly small especially at low to moderate current densities (0 to 0.25 A $cm^{-2}$). It should be noted that the impact of these parameters becomes evident at higher cell potentials where mass transport becomes dominant. Increasing anode CL and anode ECSA decreases the cell potential mainly by decreasing the anode overpotential, but results in only minimal variations in product distribution. Finally, decreasing $CO_2$ concentration (partial pressure) has a negative impact on the product of $CO_2R$ due to the competing HER at the cathode and lower reactant $CO_2$ concentration.

An important issue associated with $C_2H_4$ synthesis in an OER|AEM|$CO_2$R MEA is the crossover of $CO_2$ from the cathode to the anode chamber [22,31,32,48]. This phenomenon occurs because the AEM allows $CO_3^{2-}$ and $HCO_3^-$ anions formed in the cathode CL to cross over to the anode CL, where they reconvert to $CO_2$ due to the lower pH. Figure 7a shows the $CO_2$ utilization efficiency, defined as the fraction of $CO_2$ that is utilized effectively during the electrochemical $CO_2$R, accounting for $CO_2$ appearing in the anode flow channel and the molar flow rate of $CO_2$ that crosses over from cathode to the anode compartment, relative to the molar flow rate of $CO_2$ converted to C-containing products (Refer Eq. 17 and 57 in the section 2.5 of the Methods). For the base case, the $CO_2$ utilization efficiency is zero for cell potentials below 2.0 V and then rises to a maximum of 24.6% at 3.31 V and 0.189 A cm$^{-2}$, after which the utilization efficiency decreases with further increase in cell voltage. It is seen that $CO_2$ utilization efficiency is the highest between cell voltages of 3 to 3.5 V. All other parameters impacting the $CO_2$ utilization efficiency are compared at a current density of 0.15 A cm$^{-2}$ in Figure 7b. We observed that the highest $CO_2$ utilization efficiency (32%) is achieved by increasing cathode specific ECSA, AEM conductivity, and water transport parameters.

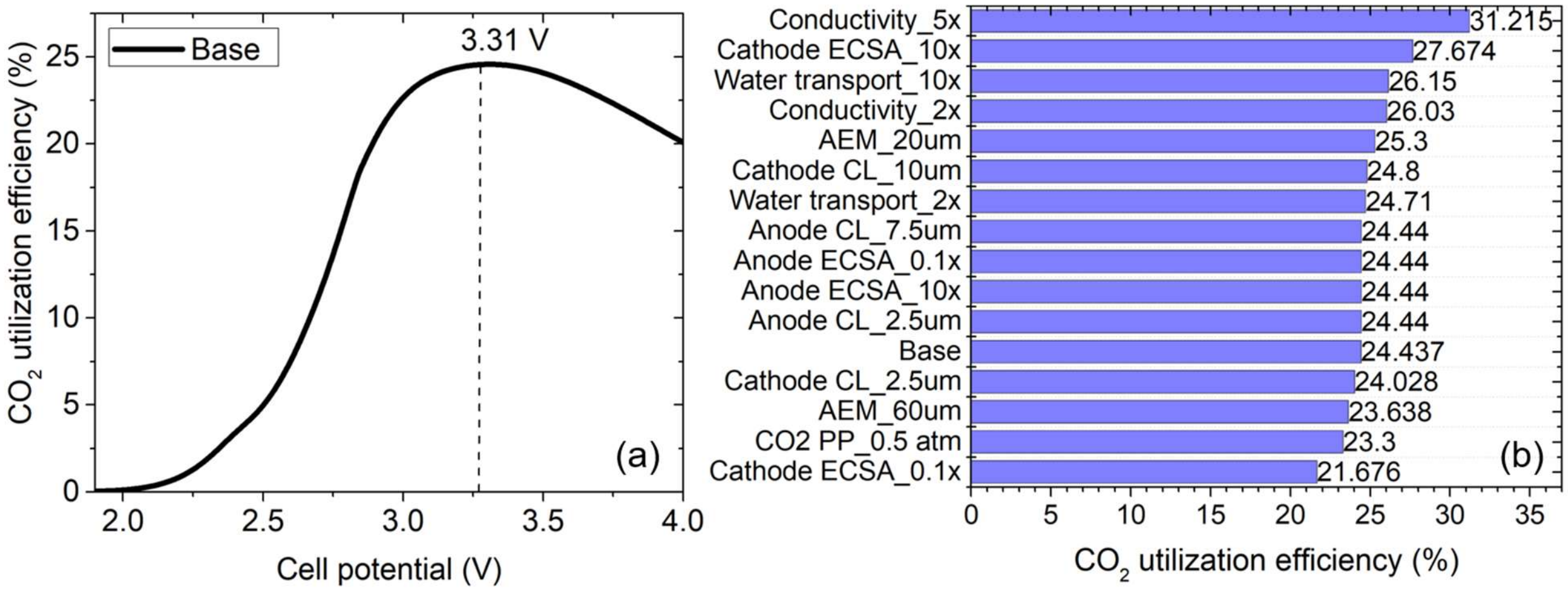

**Figure 7**. Model predictions for $CO_2$ utilization efficiency (a) as a function of cell voltage (b) for all parameters at 0.15 A cm$^{-2}$current density.

Finally, it is important to estimate the cost electricity to produce a metric tonne of $C_2H_4$ via $CO_2R$ in an MEA. The steps for this calculation are presented in section 2.7 of the methods section. Figure 8a shows the cost of electricity as a function of the cell voltage by using results for the base case. It is seen that the \$ tonne$^{-1}$ of $C_2H_4$ decreases rapidly with increasing cell voltage. This trend is consistent with the increase in FE for $C_2H_4$ with cell voltage seen in Figure 3c. If one assumes that the cost of electricity is 0.01 \$ kWh$^{-1}$, then the energy cost is 1,288 \$ tonne$^{-1}$ of $C_2H_4$ at a moderate current density of 0.15 A cm$^{-2}$. All other parameters affecting the $C_2H_4$ production cost are compared at this current density in Figure 8b. We found that the cost of electricity can be reduced to 1076 \$ tonne$^{-1}$ by reducing the cathode specific ECSA by a factor of ten. A techno-economic analysis of $C_2H_4$ synthesis in an MEA [5] projects that the cost of electricity is ~80% of the total CAPEX and OPEX for producing $C_2H_4$ by $CO_2R$, assuming that that electricity is available at 0.02 \$ kWh$^{-1}$. Taking the cost of electricity as 0.01 \$ kWh$^{-1}$ leads to a reduction of the percentage of electricity cost of the CAPEX plus OPEX to 67%, from which it can be estimated that the cost of $C_2H_4$ production rises to 1,763 \$ kWh$^{-1}$ of $C_2H_4$. This price estimate does not include the cost of separating $CO_2$ from $O_2$, produced at the anode, and recycling it to the cathode nor the cost of removing soluble products from the anolyte; however, a recent study suggests that these processes will add only 10% to the total cost of producing a tonne of $C_2H_4$ [49]. It is also noted that a part of these extra costs could be offset by the sale of ethanol (800 \$ tonne$^{-1}$) [49] and CO (200 \$ tonne$^{-1}$)[50].

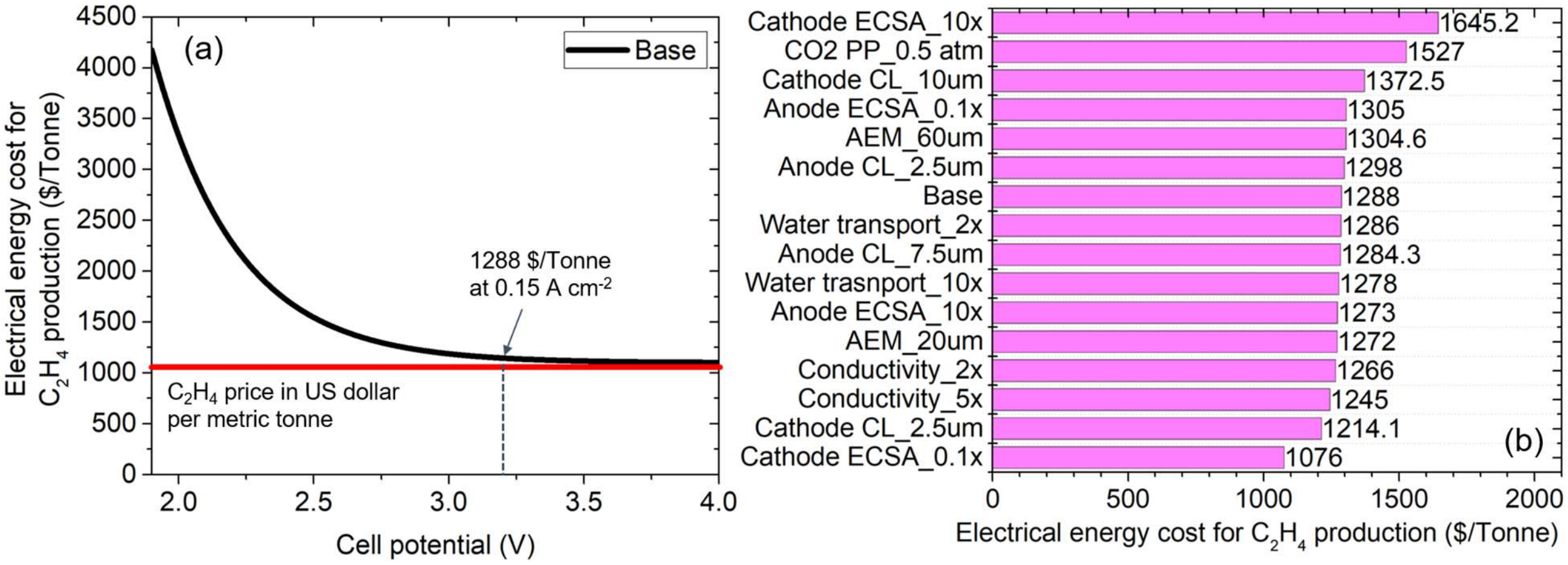

**Figure 8.** $C_2H_4$ production cost from Electrochemical $CO_2R$ (a) as a function of cell voltage (b) all parameters at 0.15 A $cm^{-2}$current density.

Since the largest expense for $C_2H_4$ production is for electricity, we explored whether that cost could be reduced by increasing the FE to this product. Our results showed that FE is enhanced at elevated cell voltages. Table 2 compares key parameters affecting electricity cost at an operating cell voltage of 3.75 V. At this voltage, we found that modifying $CO_2R$ cell components can reduce the electricity cost to \$993 per tonne of $C_2H_4$. To further improve performance, we experimentally modified the $CO_2R$ cell by coating the Cu particles in the cathode CL with Nafion, a cation-exchange membrane, but retaining the PiperION coating with the IrOx particles in the anode CL, as well as the PiperION membrane. This configuration reduced the electricity cost to 761 \$ $tonne^{-1}$ of $C_2H_4$, resulting in an estimated total production cost of 1,135 \$ $tonne^{-1}$, which is close to the current U.S. market price of ~1,000 \$ $tonne^{-1}$ of $C_2H_4$[51]. These unpublished results are presented in Table S5 in the SI. Moreover, if credit is taken for the sale of ethanol produced (0.63 tonne of $C_2H_5OH$ per tonne of $C_2H_4$), then the total cost for producing both $C_2$ products decreases to 746 \$ $tonne^{-1}$ of $C_2$. However, this price is based on the cost of electricity being 0.01 \$ $kWh^{-1}$. If the cost of electricity is 0.02 \$/kWh, then the price rises to 1,189 \$ $tonne^{-1}$ $C_2$. While this price is high, it is reasonable to expect that it could be reduced to 1,000 $ tonne $C_2$ product via reduction of the CAPEX. We also note that while the U.S. Department of Energy (DOE) estimates the cost of solar electricity to be 0.03 $/kWh in 2025 and 0.02 $/kWh in 2030 [52], there are projections that the price electricity could drop to 0.01 $/kWh sometime after 2035 [53].

**Table 2** The impact of different parameters on the cost of electricity to produce $C_2H_4$ via $CO_2R$ in an OER|AEM|$CO_2R$ MEA.

| | Cell voltage (V) | Total current density (A cm$^{-2}$) | $C_2H_4$ FE (%) | $C_2H_4$ partial current density (A cm$^{-2}$) | Electrical energy cost for $C_2H_4$ production ($ tonne$^{-1}$) |
|---|---|---|---|---|---|
| Base case | 3.75 | 0.277 | 36.4 | 0.101 | 1182 |
| AEM thickness - 20μm | 3.75 | 0.558 | 36.5 | 0.204 | 1181 |
| Cathode catalyst layer thickness - 2.5μm | 3.75 | 0.275 | 36.9 | 0.102 | 1166 |
| Cathode ECSA - 0.1x base case | 3.75 | 0.253 | 40.2 | 0.102 | 1071 |
| Conductivity - 5x base case | 3.75 | 0.319 | 43.4 | 0.138 | 993 |

# 4. Conclusions

A 1-D multiphysics model of an MEA cell for the reduction of $CO_2$ to $C_2H_4$ was developed and validated against measured cell performance and product distribution data. The validated model then was exercised to ascertain the impacts of changes of various material properties including the thickness and ECSA of the anode and cathode catalysts layers (CLs), the thickness of the AEM membrane and its ionic conductivity and water permeability, on the faradaic efficiency (FE) for $C_2H_4$ production and the cell voltage required to achieve a given current density. Among the parameters studied, cathode specific ECSA and cathode catalyst layer thickness were found to be the largest factors affecting $C_2H_4$ selectivity, energy costs for $C_2H_4$, and $CO_2$ utilization efficiency. We found that for a given current density, halving thickness of the cathode catalyst layer enhanced the FE to $C_2H_4$ by 2% and reduced the cell voltage by 40 mV. Conversely, reducing the cathode layer ECSA by a factor of ten resulted in a 7% increase in FE to $C_2H_4$, however, this also led to a 150 mV cell potential increase. Similarly, increasing the AEM thickness by $1.5 \times$ led to a 1% improvement in FE to $C_2H_4$ while introducing a 140 mV increase in cell voltage and exacerbating ion transport limitations within the cell. We also observed that the ion transport limitations within the cell and the associated cell voltage penalty can be mitigated by increasing the AEM conductivity, water permeability, as well as by reducing the AEM thickness.While a reduced cell voltage enhances $CO_2$ utilization efficiency, it also leads to a decline in the FE for $C_2H_4$ production. This tradeoff arises because local potential within the cathode catalyst layer is the primary driving force for $C_2H_4$ formation, with a 100 mV local potential increase within the cathode catalyst layer leading to 2% increase in FE to $C_2H_4$. Consequently, either an increase in the cell potential or a decrease in the FE toward $C_2H_4$ resulted in an increased electrical energy cost for producing $C_2H_4$. Finally, our cost analysis demonstrated that, with appropriate adjustments to $CO_2$R MEA design and operational parameters, it is possible to achieve $C_2H_4$ production costs that are competitive with the current market price. However, this requires the cost of electricity to decrease to 0.01 \$/kWh. Our work underscores the potential and pathway for cost-effective, sustainable $C_2H_4$ production using $CO_2$R MEAs, and provides a valuable framework for guiding future efforts aimed at optimizing efficiency and economic viability of this technology.

# Supporting Information

## S1. Kinetic Parameters

Exchange current density ($i_0$) and charge transfer coefficient ($\alpha$) are distinct but complementary kinetic parameters. The exchange current density is a measure of the intrinsic rate of the electrochemical reaction at equilibrium. It depends on electrode material, reactant concentration, and temperature. Higher $i_0$ indicates faster reaction kinetics on the catalyst surface. Thus, products having higher $i_0$ are facilitated over products having lower $i_0$. Under non-equilibrium conditions, how the applied potential drives the reaction (the symmetry of the energy barrier for electron transfer) is described by charge transfer coefficient. Products having higher charge transfer coefficients are facilitated faster than the products having lower charge transfer coefficients at higher overpotentials. Although there are very distinct kinetic parameter fit values reported in the literature, we observed similar relationship between major products (HER, COER, $C_2H_4$ ) on the Cu catalyst for the exchange current density ( $i_{CO} > i_{H_2} > i_{C_2H_4}$) and charge transfer coefficient ($\alpha_{C_2H_4} > \alpha_{H_2} > \alpha_{CO}$) [22]. We kept the same relations for exchange current density and charge transfer coefficient in our simulation to predict our experimental measurements. It should be noted that the kinetic parameter for HER is very close to the kinetic parameters for $C_2H_4$ formation. While the exchange current densities are in the same order of magnitude, charge-transfer coefficients are 0.46 and 0.44 for $C_2H_4$ formation and HER respectively. Thus, HER is facilitated at higher potentials even if there is high pH locally.

**Table S3** Rate parameters for charge-transfer reactions [30].

| | $U_k^0(V\ vs\ SHE$ $(PH=0)$ | $i_{0,k}(mA\ cm^{-2})$ | $\alpha_{a/c,k}$ | $\prod_j a_j^{\gamma_{j,k}}$ |
|---|---|---|---|---|
| $O_2$(acid) | 1.23 | $9.4x10^{-7}\exp\left(-\frac{(11+pH)\left[\frac{kJ}{mol}\right]}{RT}\right)$ | 1.5 | $a_w$ |
| $O_2$(base) | 1.23 | $1.23x10^{-4}\exp\left(-\frac{(11+pH)\left[\frac{kJ}{mol}\right]}{RT}\right)$ | 1.5 | $\frac{c_{OH^-}}{1M}$ |
| $H_2$ | 0 | $5x10^{-8}\exp\left(-\frac{(1+pH)\left[\frac{kJ}{mol}\right]}{RT}\right)$ | 0.44 | $a_w$ |
| $HCOO^-$ | –0.07 | $8x10^{-6}$ | 0.33 | $a_w\frac{c_{CO_2}}{0.034M}$ |
| CO | –0.06 | $2x10^{-6}$ | 0.36 | |
| $C_2H_4$ | 0.09 | $8x10^{-9}$ | 0.46 | $a_w\frac{c_{CO_2}}{0.034M}$ |
| $C_2H_5OH$ | 0.10 | $1x10^{-8}$ | 0.43 | $a_w\frac{c_{CO_2}}{0.034M}$ |
| $C_3H_7OH$ | 0.11 | $5x10^{-9}$ | 0.42 | $a_w\frac{c_{CO_2}}{0.034M}$ |

**Table S4** Equilibrium constants and forward rate constants of disassociation reaction at 298 K [22].

| Reaction | Constant | Units |
|---|---|---|
| $K1$ | $4.27\times10^{-7}$ | $M$ |
| $k1$ | $3.71\times10^{-2}$ | $M\ s^{-1}$ |
| $K2$ | $4.58\times10^{-11}$ | $M$ |
| $k2$ | $59.44$ | $M\ s^{-1}$ |
| $K3$ | $4.27\times10^{7}$ | $M$ |
| $k3$ | $2.23\times10^{3}$ | $M\ s^{-1}$ |
| $K4$ | $4.58\times10^{3}$ | $M$ |
| $k4$ | $6.0\times10^{9}$ | $M\ s^{-1}$ |
| $K5$ | $1\ \times\ 10^{-14}$ | $M$ |
| $k5$ | $8.9\times10^{-4}$ | $M\ s^{-1}$ |
| $K6$ | $2.05\ \times\ 10^{-4}$ | $M$ |
| $k6$ | $4.0\times10^{5}$ | $M\ s^{-1}$ |

## S2. Additional Model Parameters

**Table S5** Intrinsic model parameters

| Parameter | Value | Unit | Description | Reference |
|---|---|---|---|---|
| $\varepsilon_{DL}^{0}$ | 0.8 | | DL porosity | [28,54] |
| $\varepsilon_{CL}^{0}$ | 0.5 | | CL porosity | [28,30] |
| $f_{I,DL}$ | 0 | | Ionomer volume fraction in DL pore space | [30] |
| $f_{I,CL}$ | 0.4 | | Ionomer volume fraction in CL pore space | [30] |
| $\sigma_{s,DL}$ | 220 | $S\ m^{-1}$ | DL electronic conductivity | [28,30,54] |
| $\sigma_{s,CL}$ | 100 | $S\ m^{-1}$ | CL electronic conductivity | [28,30,55] |
| $K_{DL}$ | $8.4 \times 10^{-13}$ | $m^{-1}$ | DL Saturated permeability | [28,30,56] |
| $K_{CL}$ | $8.0 \times 10^{-16}$ | $m^{-1}$ | CL saturated permeability | [28,30,56] |
| $r_p$ | $5.0 \times 10^{-8}$ | m | Average CL particle radius | [28,30,57] |
| $D_{CO_2}$ | $2.17 \times 10^{-9}$ | $m^2 s^{-1}$ | Diffusion coefficient of $CO_2$ in water | [58] |
| $D_{H_2}$ | $4.5 \times 10^{-9}$ | $m^2 s^{-1}$ | Diffusion coefficient of $H_2$ in water | [58] |
| $D_{O_2}$ | $2.1 \times 10^{-9}$ | $m^2 s^{-1}$ | Diffusion coefficient of $O_2$ in water | [58] |
| $D_{CO}$ | $2.03 \times 10^{-9}$ | $m^2 s^{-1}$ | Diffusion coefficient of $CO$ in water | [58] |
| $D_{N_2}$ | $1.88 \times 10^{-9}$ | $m^2 s^{-1}$ | Diffusion coefficient of $N_2$ in water | [58] |
| $D_{C_2H_4}$ | $1.87 \times 10^{-9}$ | $m^2 s^{-1}$ | Diffusion coefficient of $C_2H_4$ in water | [58] |
| $H_{CO_2}$ | 34 | $mM\ atm^{-1}$ | The solubility of $CO_2$ in water | [59] |
| $H_{H_2}$ | 0.78 | $mM\ atm^{-1}$ | The solubility of $H_2$ in water | [59] |
| $H_{O_2}$ | 1.3 | $mM\ atm^{-1}$ | The solubility of $O_2$ in water | [59] |
| $H_{CO}$ | 0.95 | $mM\ atm^{-1}$ | The solubility of $CO$ in water | [59] |
| $H_{C_2H_4}$ | 4.8 | $mM\ atm^{-1}$ | The solubility of $C_2H_4$ in water | [59] |
| $H_{N_2}$ | 0.65 | $mM\ atm^{-1}$ | The solubility of $N_2$ in water | [59] |
| $k_{T,DL}$ | 0.015 | $W\ cm^{-1} K^{-1}$ | DL thermal conductivity | [29,30,60] |
| $k_{T,CL}$ | 0.003 | $W\ cm^{-1} K^{-1}$ | CL thermal conductivity | [29,30,60] |
| $k_{T,AEM}$ | 0.0025 | $W\ cm^{-1} K^{-1}$ | AEM thermal conductivity | [29,30] |
| $c_{p,DL}$ | 1000 | $J\ Kg^{-1} K^{-1}$ | DL specific heat | [29,30] |
| $c_{p,CL}$ | 2000 | $J\ Kg^{-1}K^{-1}$ | CL specific heat | [29,30] |
| $c_{p,AEM}$ | 4000 | $J\ Kg^{-1}K^{-1}$ | AEM specific heat | [29,30] |
| $\rho_{DL}$ | 300 | $kg\ m^{-3}$ | DL density | [29,30] |
| $\rho_{CL}$ | 500 | $kg\ m^{-3}$ | CL density | [29,30] |
| $\rho_{AEM}$ | 1200 | $kg\ m^{-3}$ | AEM density | [29,30,61] |
| $\Delta H_1$ | 7.64 | $kJ\ mol^{-1}$ | Enthalpy change of homogeneous reaction 1 | [29,30] |
| $\Delta H_2$ | 14.9 | $kJ\ mol^{-1}$ | Enthalpy change of homogeneous reaction 2 | [29,30] |
| $\Delta H_3$ | -48.2 | $kJ\ mol^{-1}$ | Enthalpy change of homogeneous reaction 3 | [29,30] |
| $\Delta H_4$ | -41 | $kJ\ mol^{-1}$ | Enthalpy change of homogeneous reaction 4 | [29,30] |
| $\Delta H_w$ | 55.8 | $kJ\ mol^{-1}$ | Enthalpy change of homogeneous water splitting reaction | [29,30] |
| $\Pi_{H_2}$ | 13 | mV | Peltier Coefficient for the $H_2$ evolution Reaction. | [29,30] |
| $\Pi_{O_2}$ | 240 | mV | Peltier Coefficient for the $O_2$ evolution Reaction. | [29,30] |
| $\Pi_{CO}$ | 38 | mV | Peltier Coefficient for the $CO$ evolution Reaction. | [29,30] |
| $\Pi_{HCOO}$ | -104 | mV | Peltier Coefficient for the $HCOO$ evolution Reaction. | [29,30] |
| $\Pi_{C_2H_4}$ | -123 | mV | Peltier Coefficient for the $C_2H_4$ evolution Reaction. | [29,30] |
| $\Pi_{C_2H_5OH}$ | -123 | mV | Peltier Coefficient for the $C_2H_5OH$ evolution Reaction. | [29,30] |
| $\Pi_{C_3H_7OH}$ | -135 | mV | Peltier Coefficient for the $C_3H_7OH$ evolution Reaction. | [29,30] |

**Table S6** Transport correlations in the Membrane/ionomer [30].

| | | |
|---|---|---|
| Water Trnasport coefficient | $a_V$ @298 K | $8.0x10^{-14}exp\ (11.47a_0)$ |
| $(mol^2J^{-1}cm^{-1}\ s^{-1})$ | $a_V$ @350 K | $2.3x10^{-13}exp\ (11.47a_0)$ |
| | $a_L$ | $100a_V^{max})$ |
| Membrane/ionomer conductivity | $K_V$ @ 298 K | $0.003exp\ (8.14a_0)$ |
| $(S\ m^{-1})$ | $K_V$ @ 350 K | $0.006exp\ (6.21a_0$ |
| | $K_L$ | $2K_L$ |
| Electro − osmotic drag coefficient | $\xi_V$ | 1.83 |
| | $\xi_L$ | 9 |
| $CO_2$solubility $(mM\ atm^{-1})$ | $H_{CO_2}$ | $34exp\left(-2400\left(\frac{1}{T[K]}-\frac{1}{298}\right)\right)$ |
| $CO_2$ diffusivity $(m^2s^{-1})$ | $D_{CO_2}$ | $2.17x10^{-9}exp\left(-2345\left(\frac{1}{T[K]}-\frac{1}{298}\right)\right)$ |

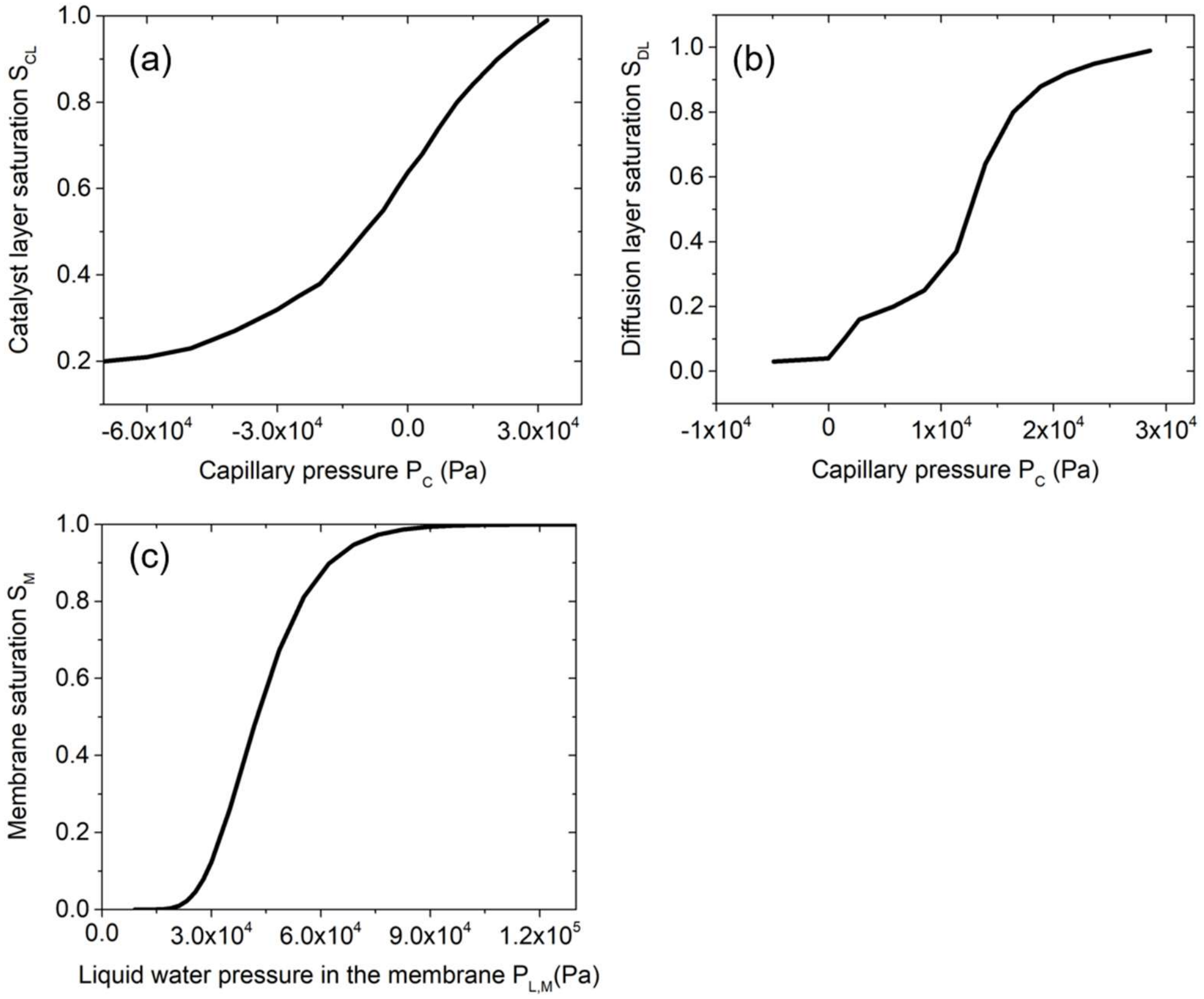

**Figure S9.** Saturation curve for (a) catalyst layer (b) diffusion layer (c) membrane to describe hydration between vapor and liquid-equilibrated states.

## S3. Applied Voltage Breakdown Analysis

The total cell voltage was decomposed into individual overpotentials. The Equilibrium potential for OER at the anode,

$$\Delta V_{Eq,OER.acid} = \int_{aCL} \frac{i_{OER.acid}(U_{OER,acid}^{0} - \frac{2.303RT}{F}\text{pH})}{i_{total}}\,\text{dx} \tag{S68}$$

$$\Delta V_{Eq,OER.base} = \int_{aCL} \frac{i_{OER,base}(U_{OER,base}^{0} - \frac{2.303RT}{F}\text{pH})}{i_{total}}\,\text{dx} \tag{S69}$$

The Equilibrium potential of the HER, and $CO_2R$ reactions are calculated as follows,

$$\Delta V_{Eq,k} = \sum_{k} \int_{cCL} \frac{i_{k}(U_{k}^{0} - \frac{2.303RT}{F}\text{pH})}{i_{total}}\,\text{dx} \tag{S70}$$

The overpotentials associated with OER at the anode,

$$\Delta V_{OER,acid} = \int_{aCL} \frac{\eta_{OER,acid} i_{OER,acid}}{i_{total}}\,\text{dx} \tag{S71}$$

$$\Delta V_{OER,base} = \int_{aCL} \frac{\eta_{OER,base} i_{OER,base}}{i_{total}}\,\text{dx} \tag{S72}$$

The overpotentials associated with HER, and $CO_2R$ reactions at the cathode,

$$\Delta V_{k} = \sum_{k} \int_{cCL} \frac{\eta_{k} i_{k}}{i_{total}}\,\text{dx} \tag{S73}$$

The ohmic overpotential due to ion transport,

$$\Delta V_{ionic} = \int_{aCL+AEM+cCL} \frac{\boldsymbol{i_l} \cdot \nabla \phi_l}{i_{total}} \, dx \tag{S74}$$

The ohmic overpotential due to the electron transport,

$$\Delta V_{electrical} = \int_{aCL+aGDL+cCL+cGDL} \frac{\boldsymbol{i_s} \cdot \nabla \phi_s}{i_{total}} \, dx \tag{S75}$$

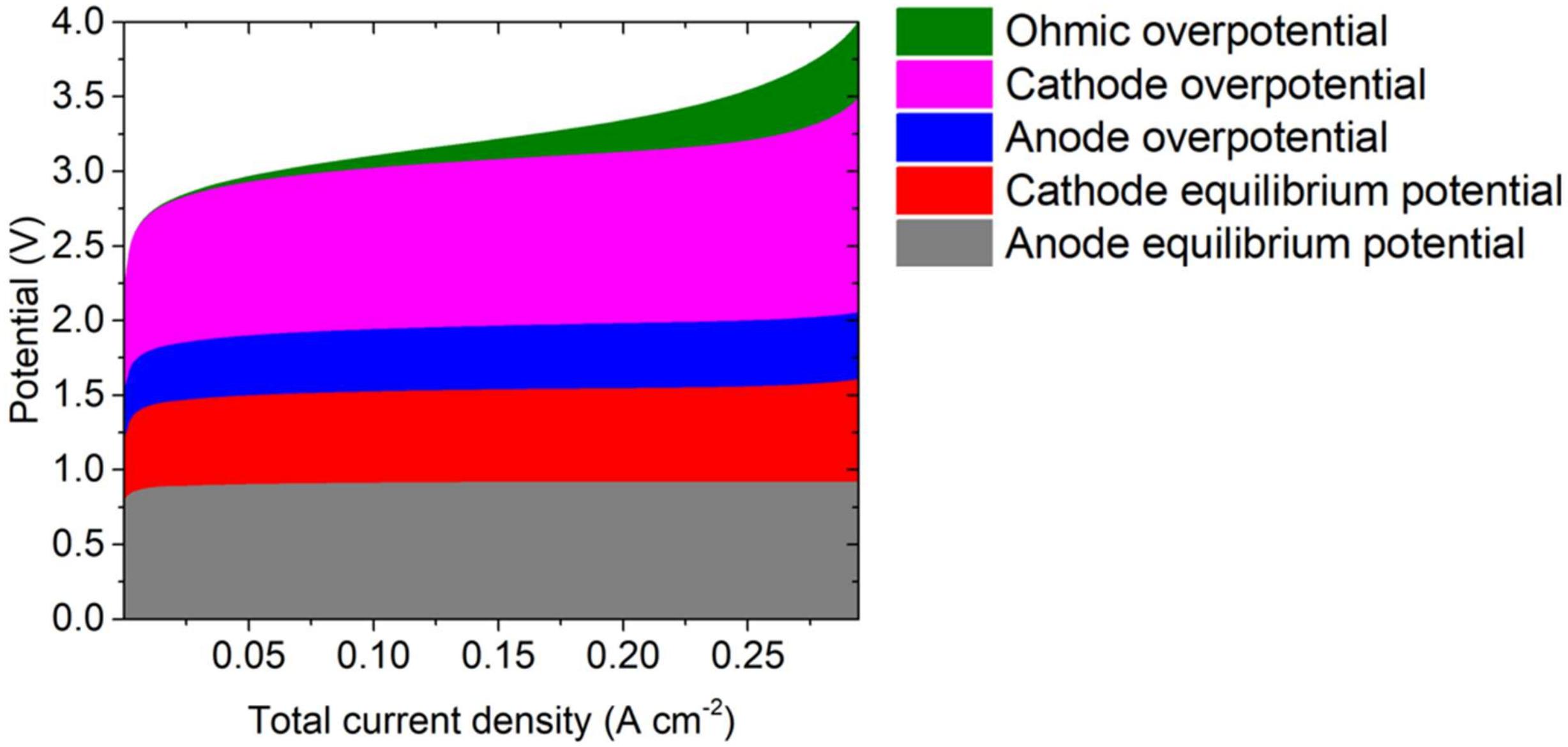

**Figure S10.** The model predictions for the base case applied-voltage breakdown.

## S4. Supplementary Figures

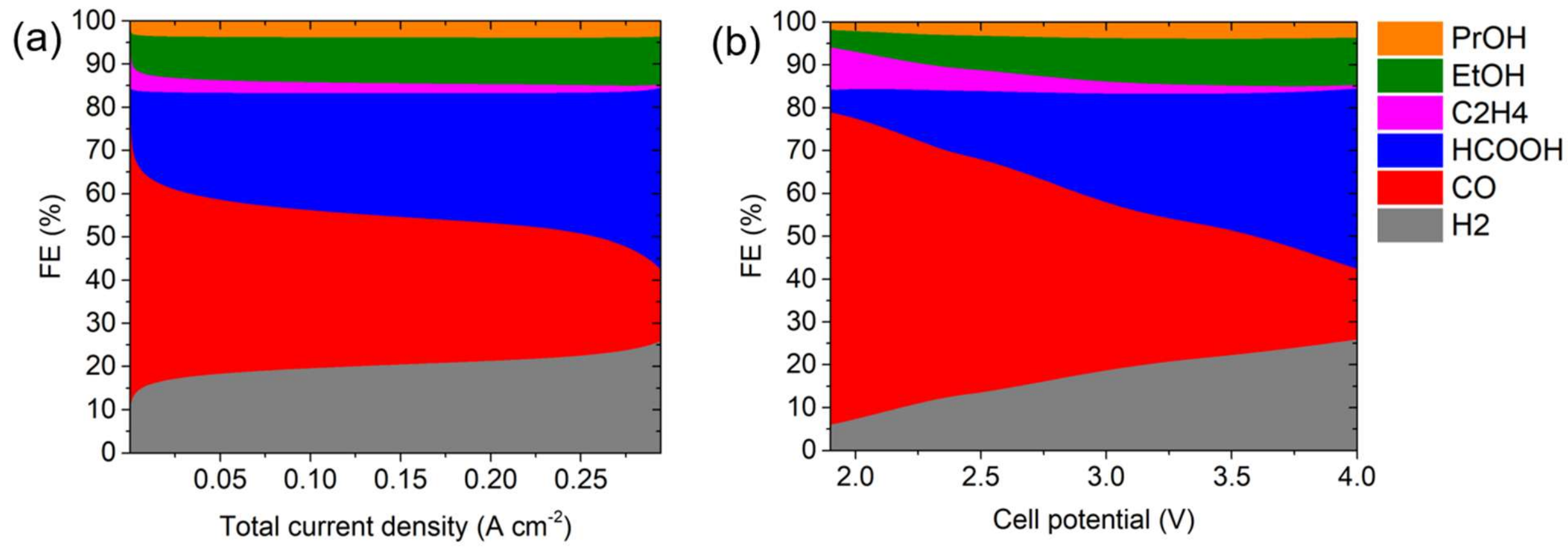

**Figure S11** The model predictions for the base case (a) product distribution as a function of current density (b) product distribution as a function of cell potential.

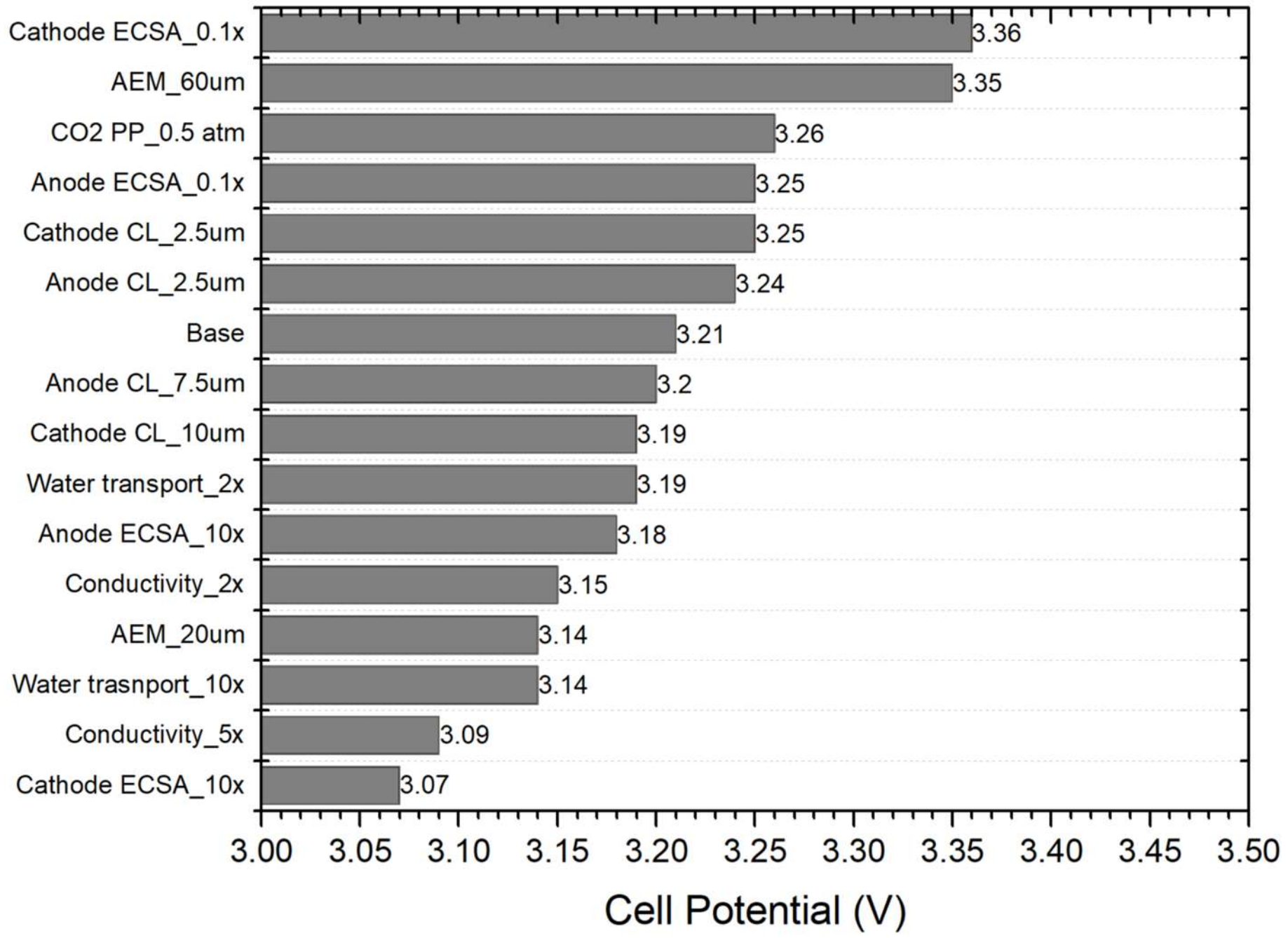

**Figure S12.** Predicted cell potentials for all parameters at 0.15 A $cm^{-2}$current density.

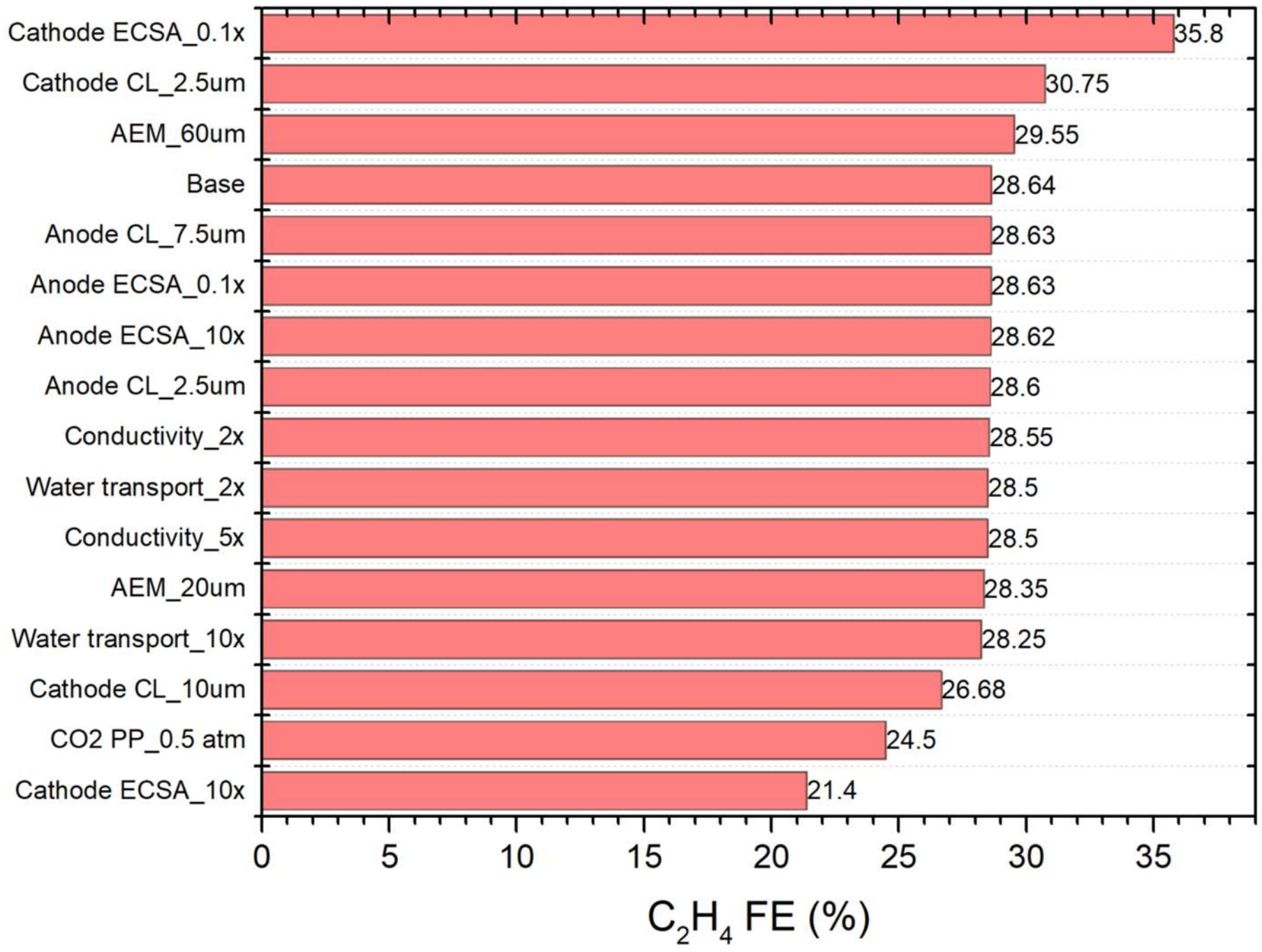

**Figure S13.** Predicted FEs for all parameters at 0.15 A $cm^{-2}$ current density.

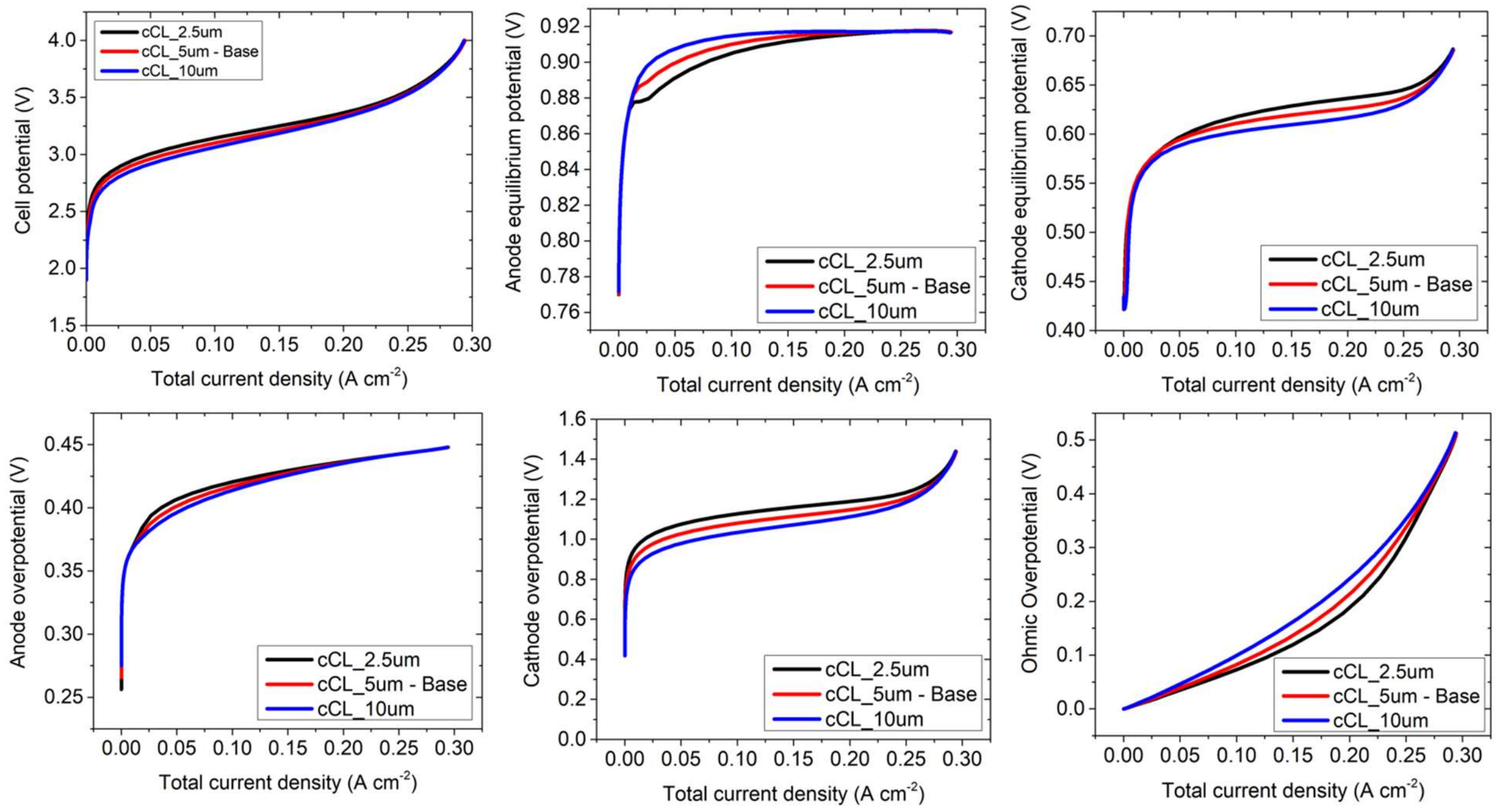

**Figure S14.** Polarization curve and individual overpotential contributions as a function of cathode catalyst layer thickness.

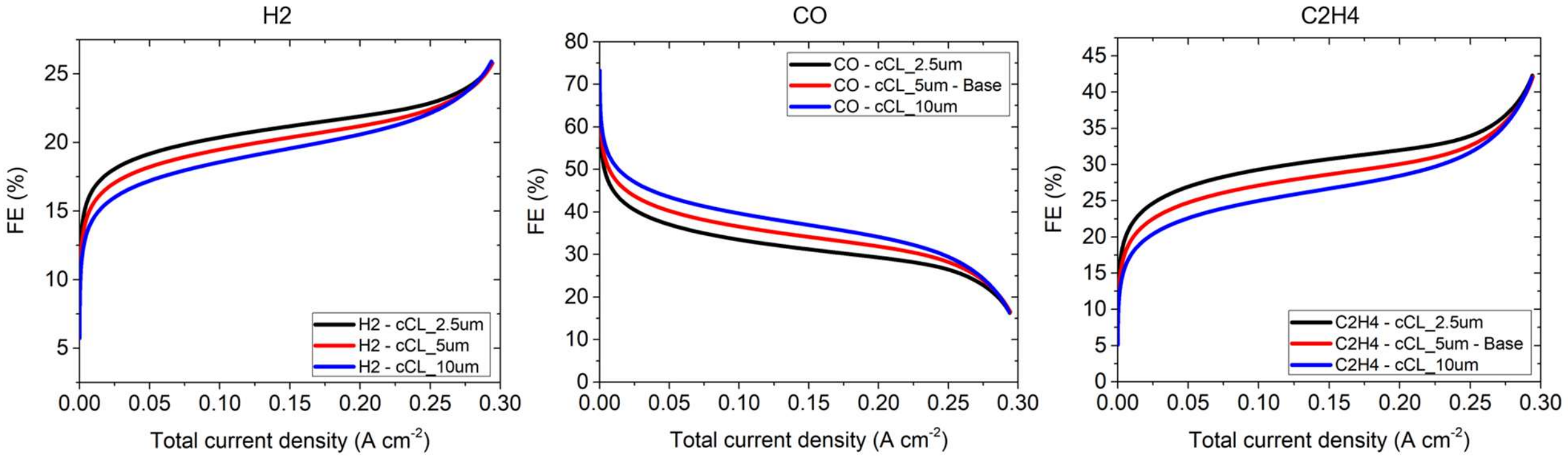

**Figure S15.** Faradaic efficiencies for $H_2$ and major $CO_2R$ products as a function of cathode catalyst layer thickness.

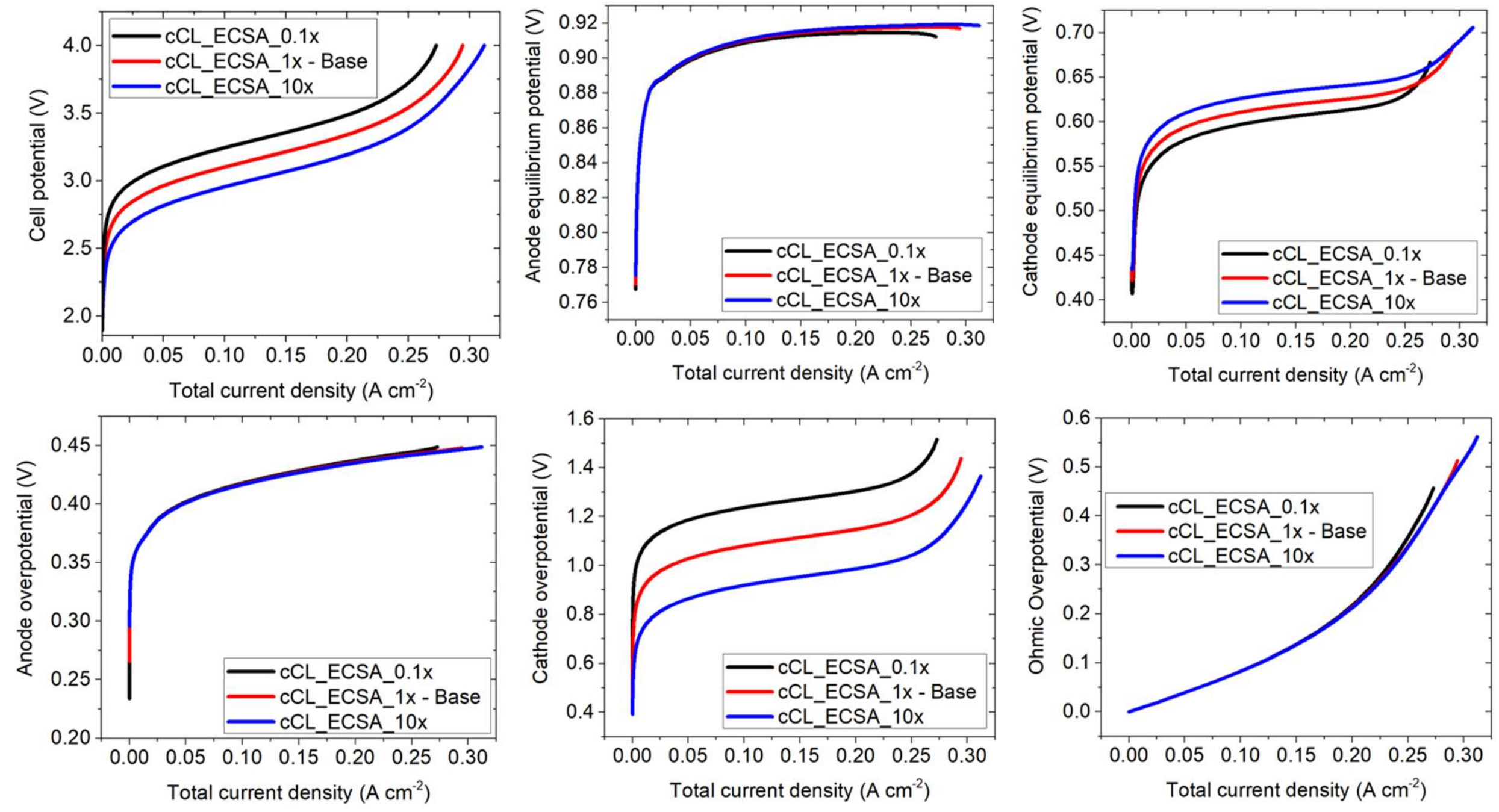

**Figure S16.** Polarization curve and individual overpotential contributions as a function of cathode ECSA.

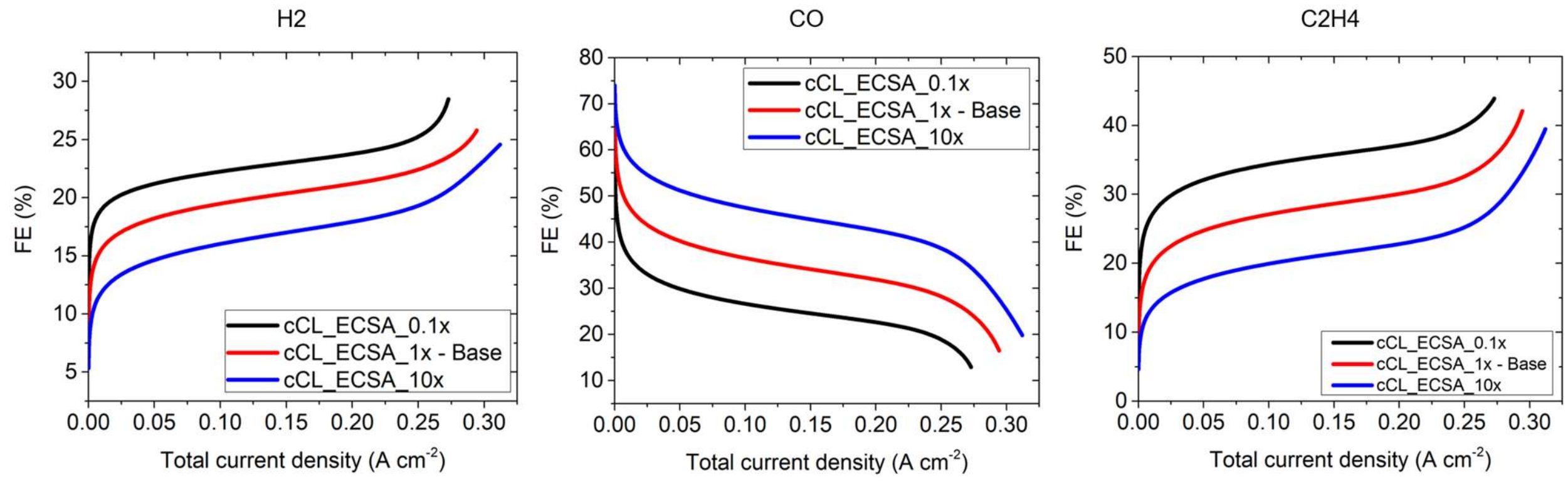

**Figure S17.** Faradaic efficiencies for $H_2$ and major $CO_2R$ products as a function of cathode ECSA.

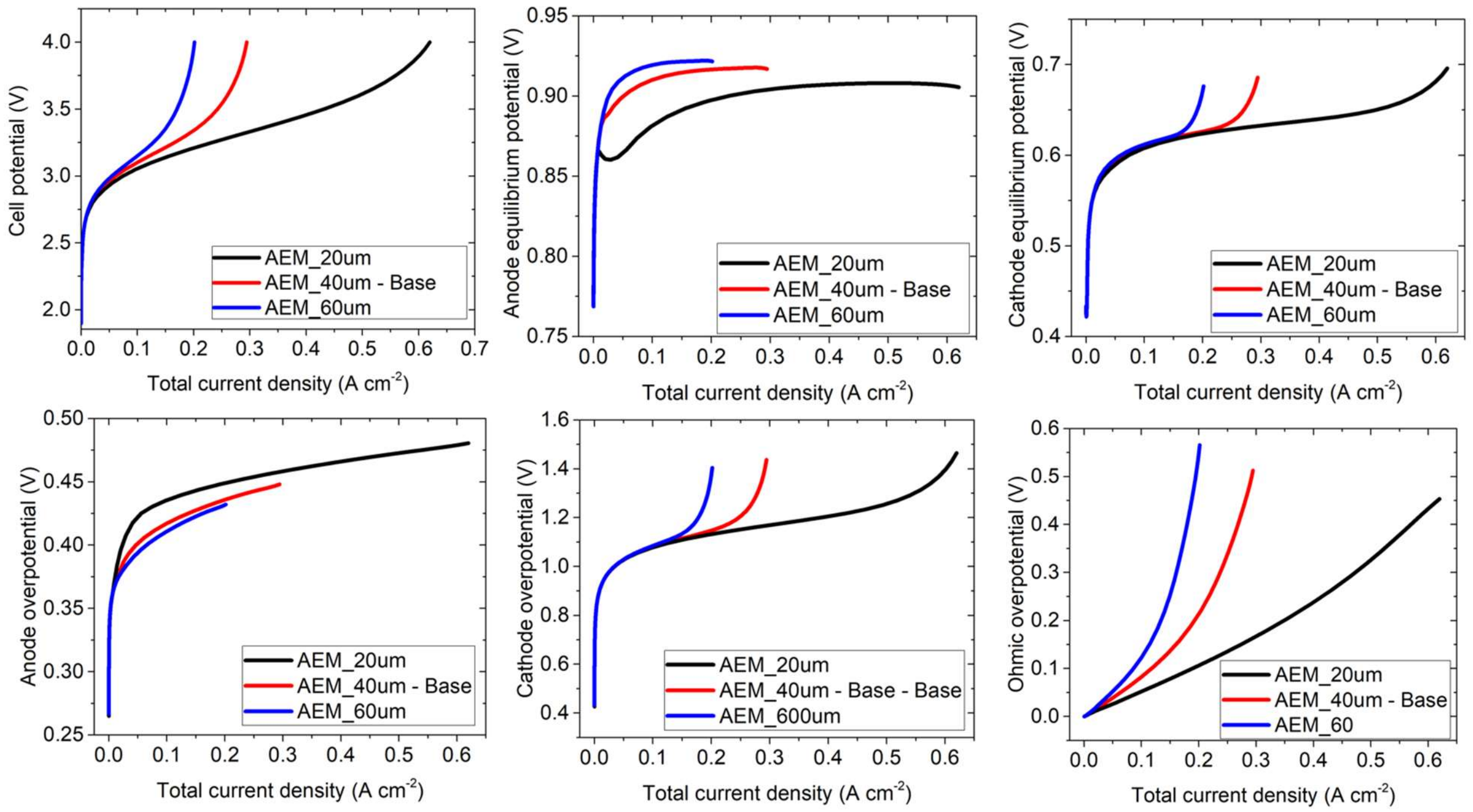

**Figure S18.** Polarization curve and individual overpotential contributions as a function of AEM thickness.

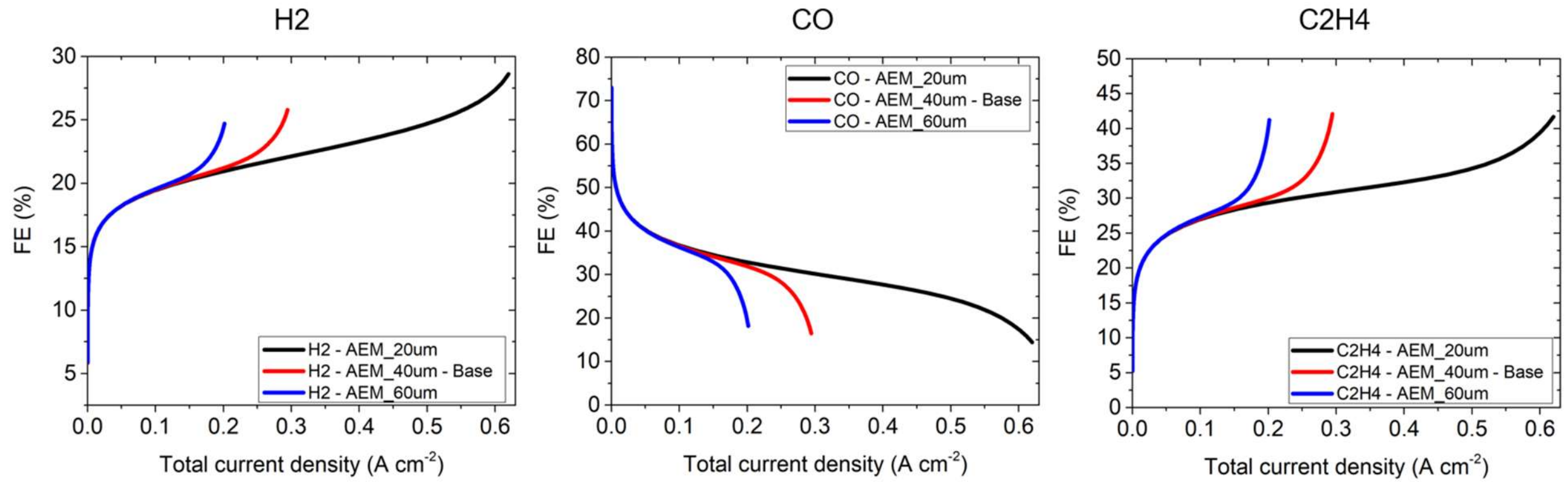

**Figure S19.** Faradaic efficiencies for $H_2$ and major $CO_2R$ products as a function of AEM thickness.

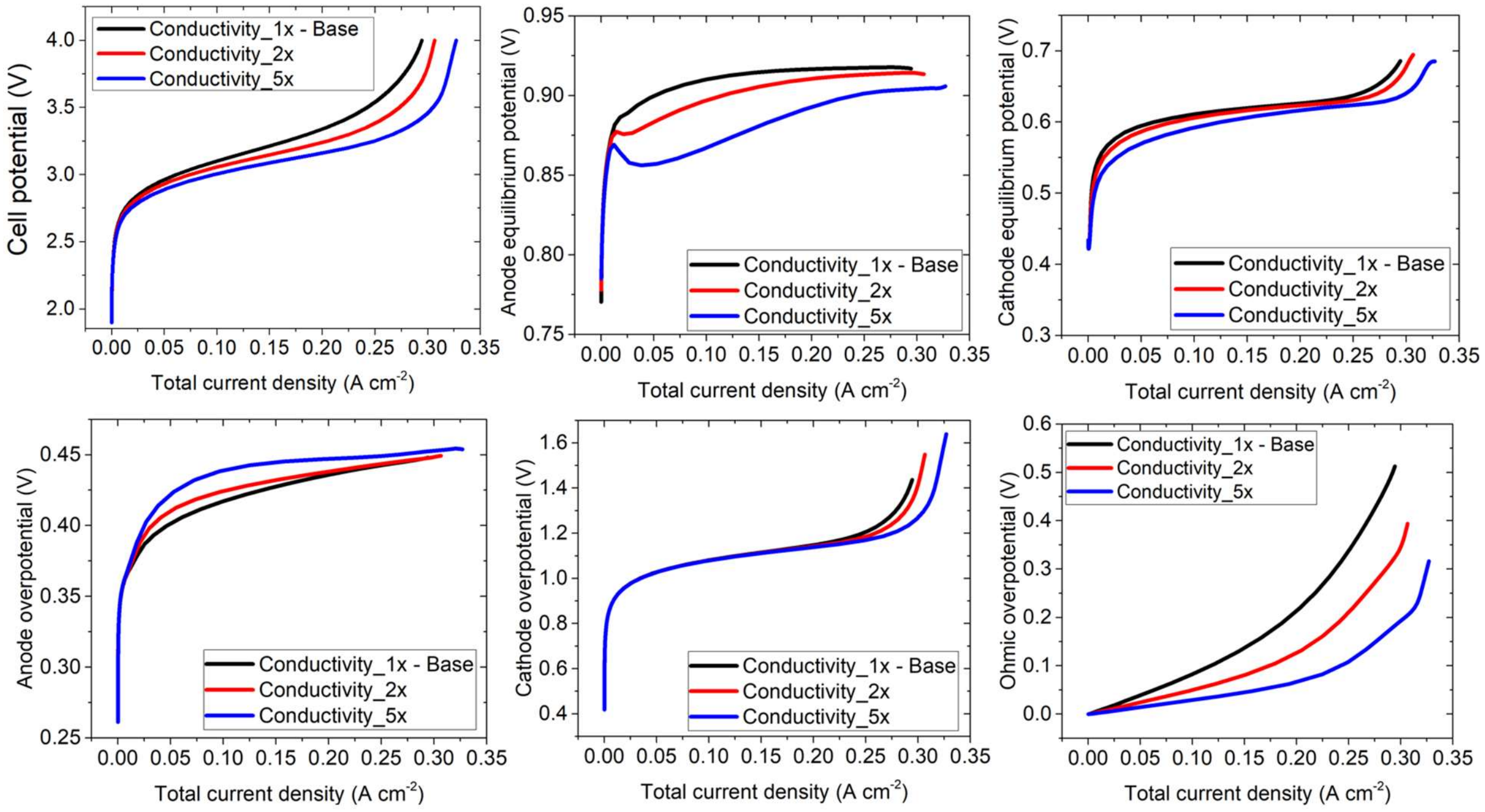

**Figure S20.** Polarization curve and individual overpotential contributions as a function of conductivity.

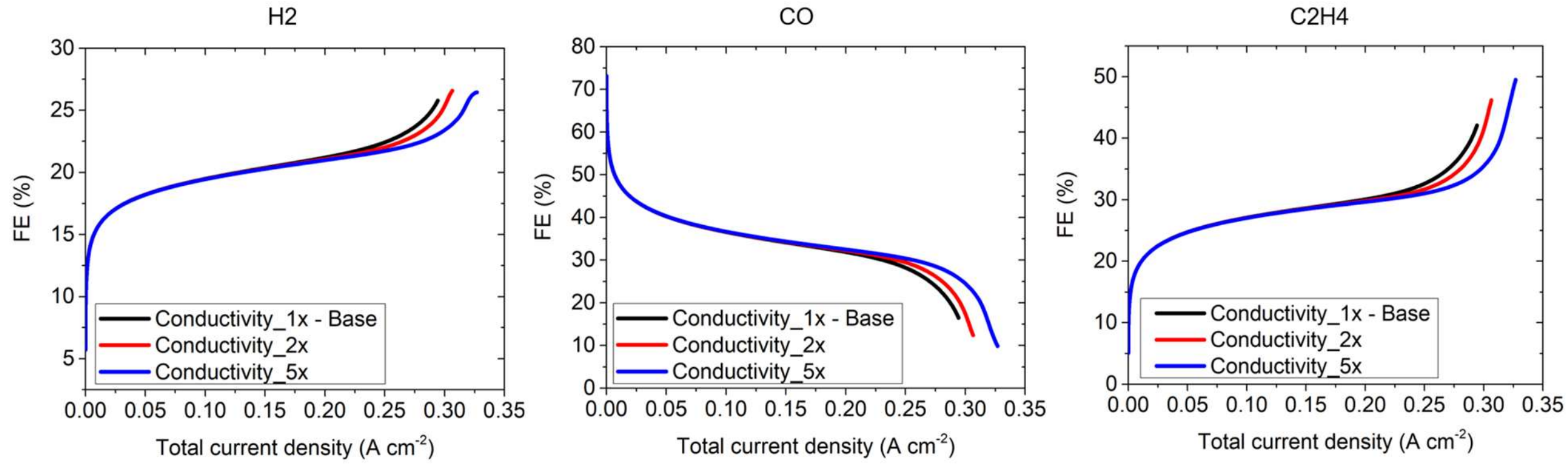

**Figure S21.** Faradaic efficiencies for $H_2$ and major $CO_2R$ products as a function of conductivity.

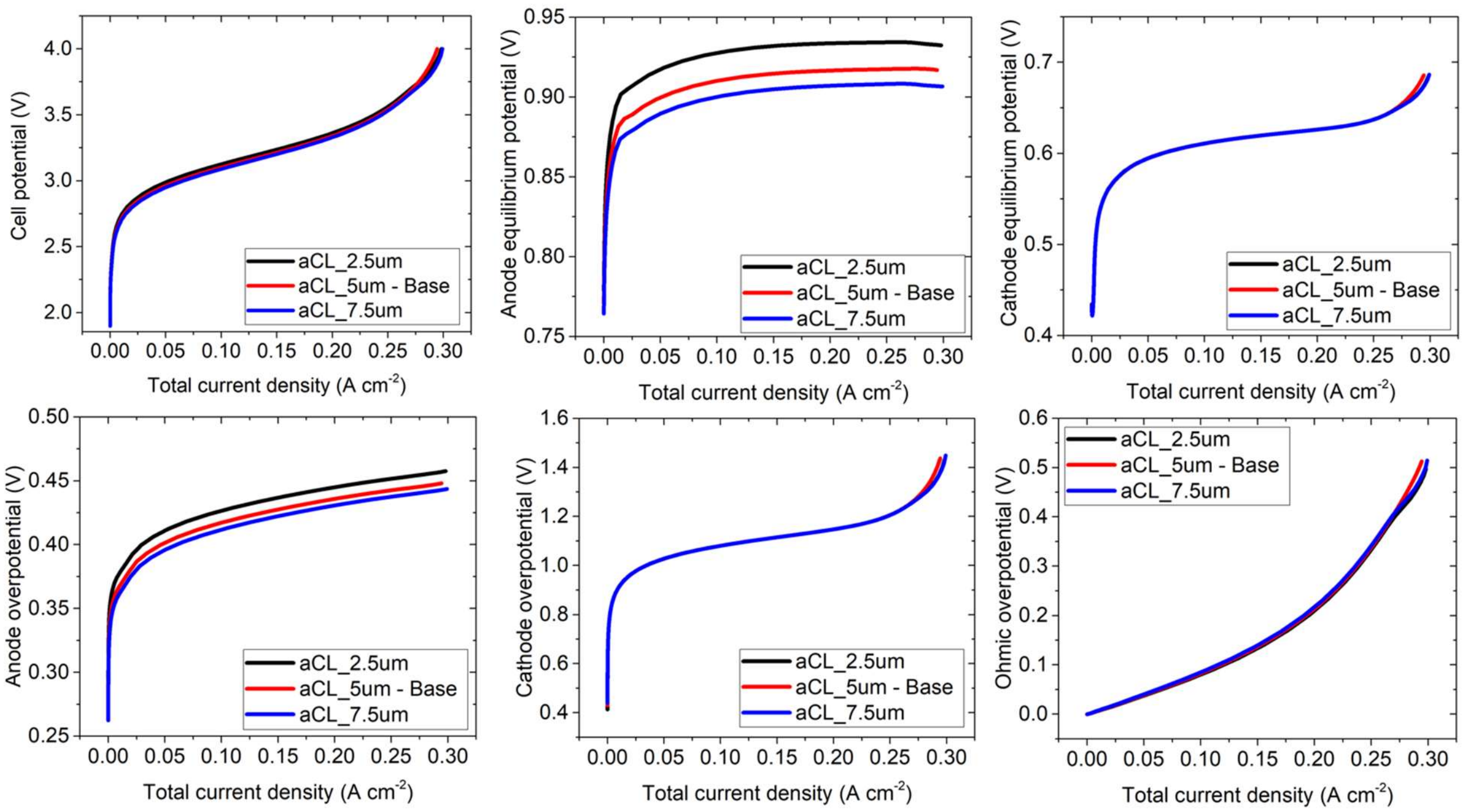

**Figure S22.** Polarization curve and individual overpotential contributions as a function of anode catalyst layer thickness.

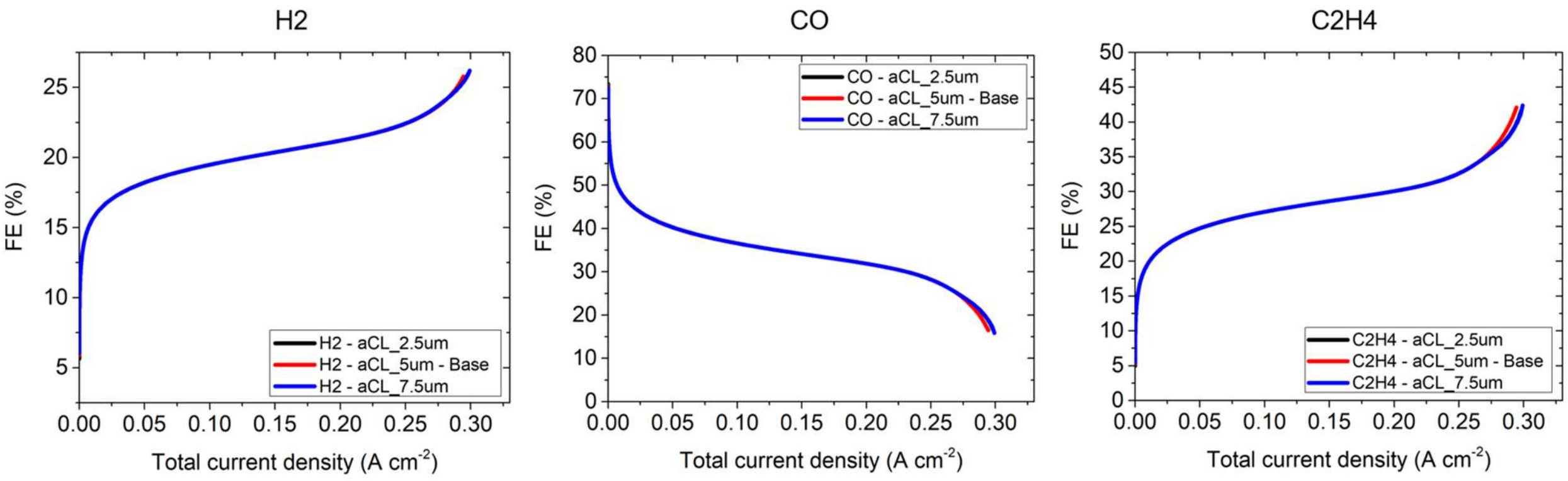

**Figure S23.** Faradaic efficiencies for $H_2$ and major $CO_2R$ products as a function of anode catalyst layer thickness.

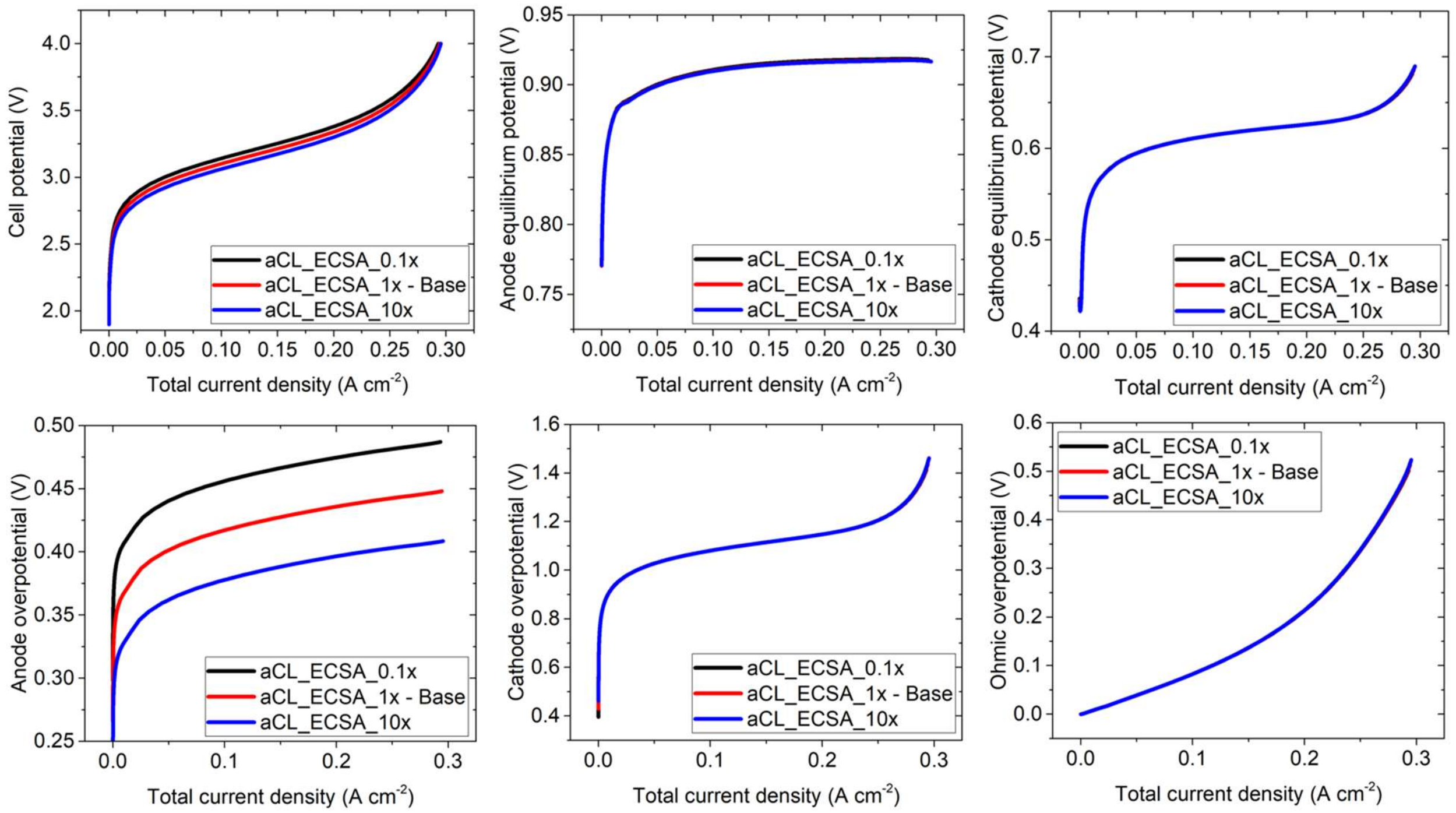

**Figure S24.** Polarization curve and individual overpotential contributions as a function of anode ECSA.

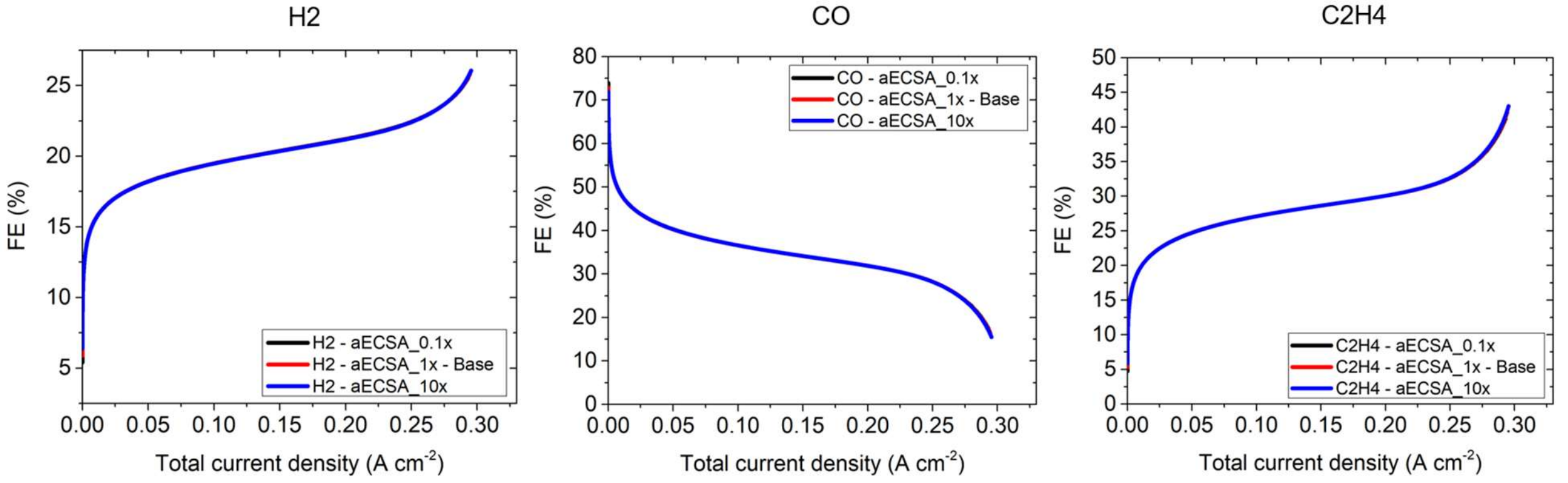

**Figure S25.** Faradaic efficiencies for $H_2$ and major $CO_2R$ products as a function of anode ECSA.

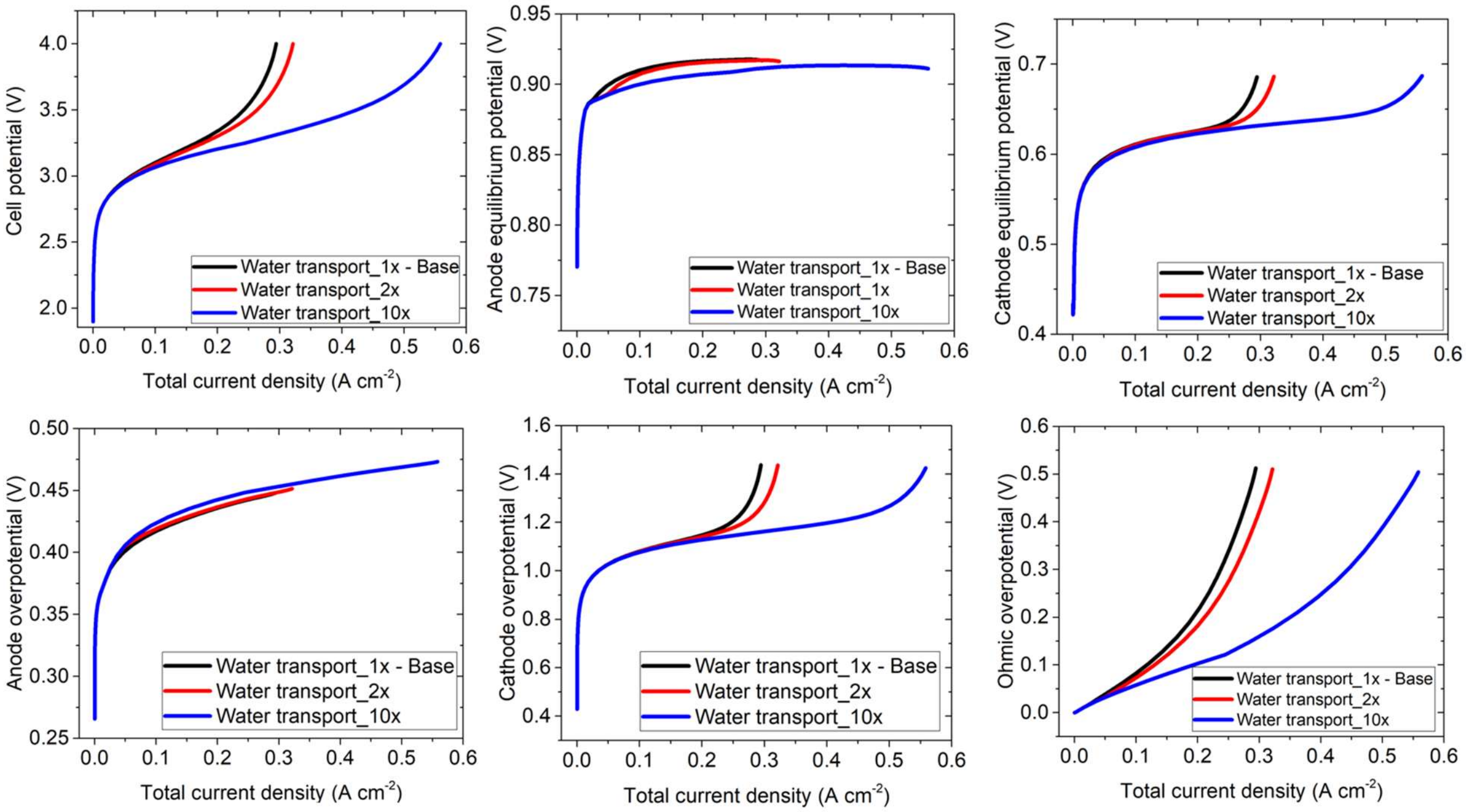

**Figure S26.** Polarization curve and individual overpotential contributions as a function of water transport.

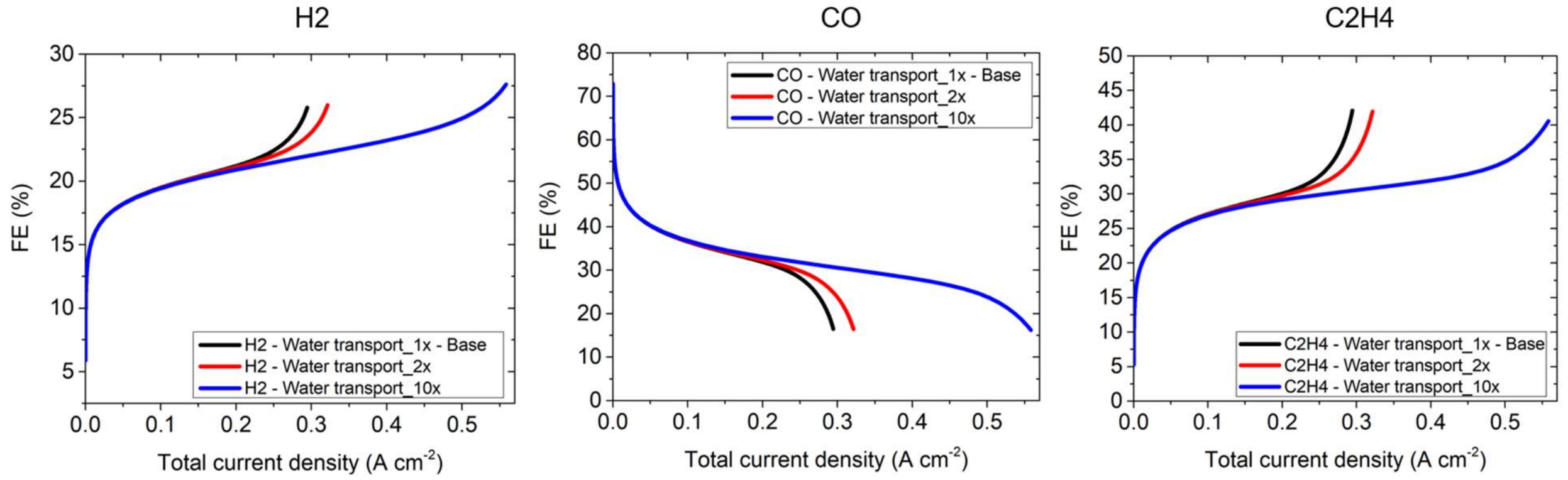

**Figure S27.** Faradaic efficiencies for $H_2$ and major $CO_2R$ products as a function of water transport.

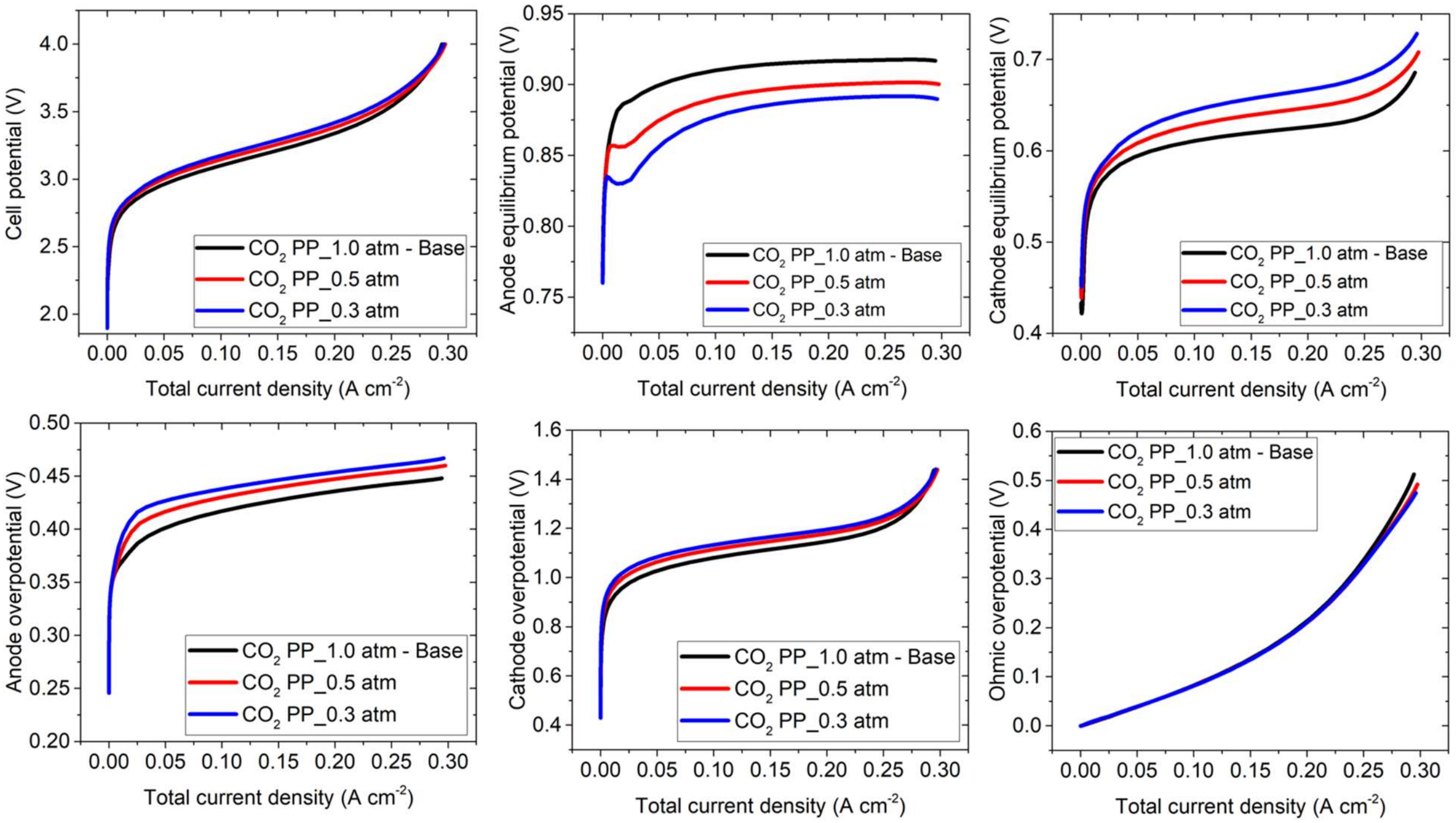

**Figure S28.** Polarization curve and individual overpotential contributions as a function of $CO_2$ partial pressure.

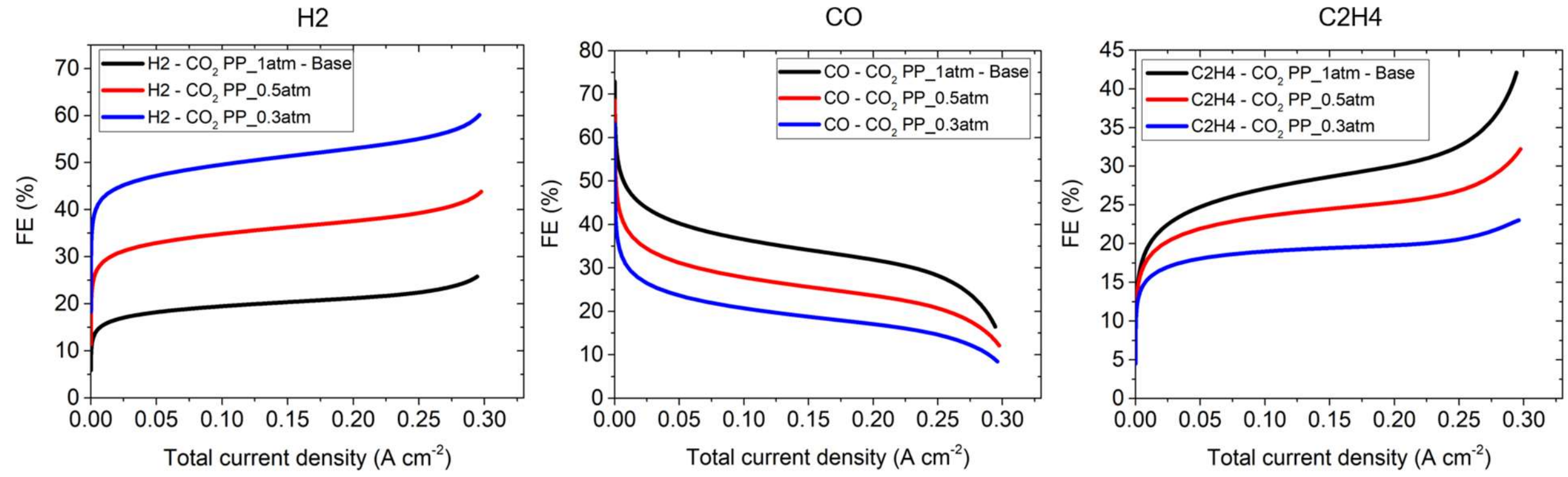

**Figure S29.** Faradaic efficiencies for $H_2$ and $CO_2$R products as a function of CO2 partial pressure.

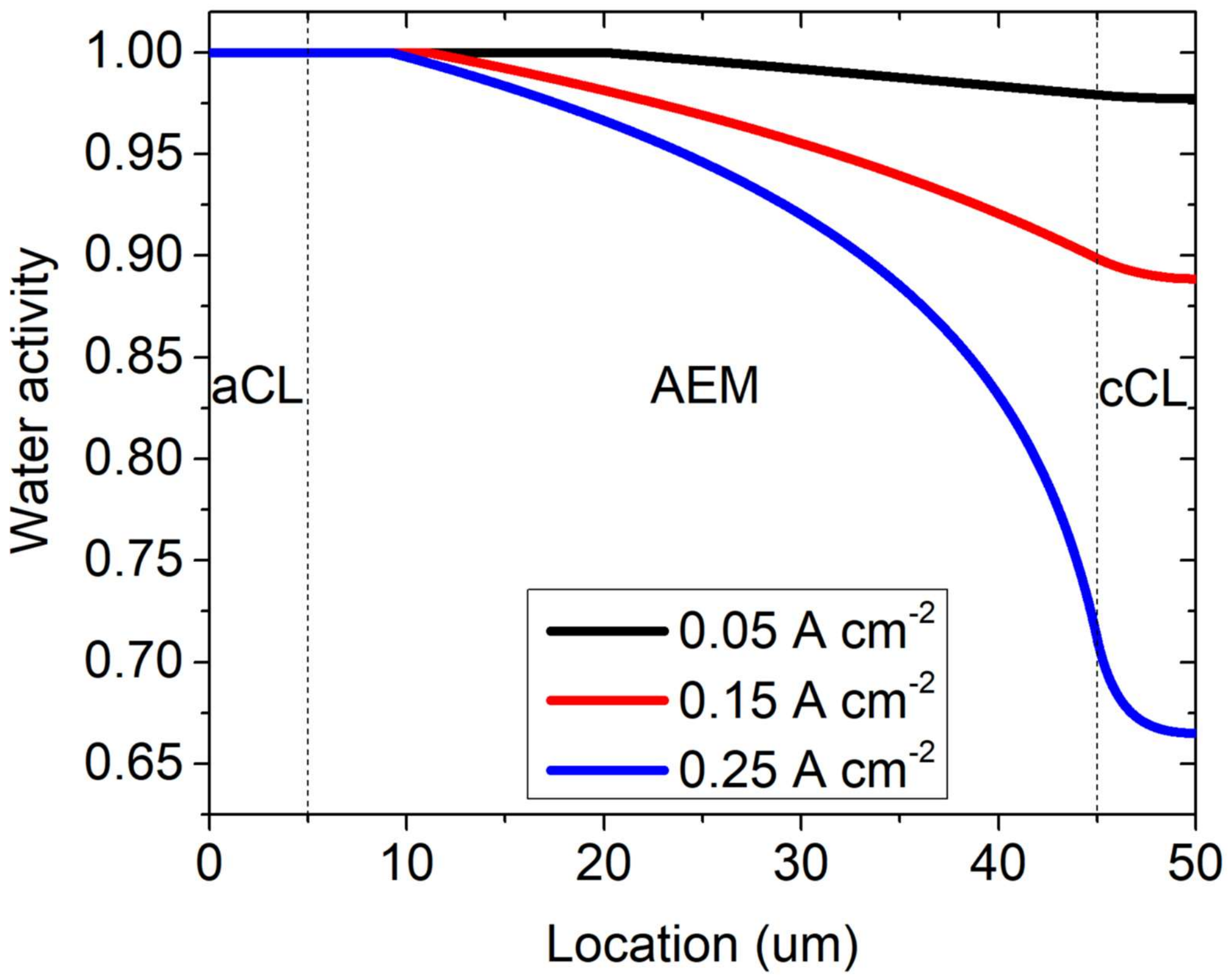

**Figure S30.** Electrical energy cost for $C_2H_4$ production via electrochemical $CO_2R$ in MEA at 0.15 A cm$^{-2}$ current density.

**Table 7** The impact of experimental modifications to the $CO_2R$ cell on the cost of electricity to produce $C_2H_4$ via $CO_2R$ in an OER|AEM|$CO_2R$ MEA.

| | Cell voltage (V) | Total current density (A $cm^{-2}$) | $C_2H_4$ FE (%) | $C_2H_4$ partial current density (A $cm^{-2}$) | Electrical energy cost for $C_2H_4$ production (\$ $tonne^{-1}$) |
|---|---|---|---|---|---|
| Improved cCL layer (see text) | 3.75 | 0.438 | 58.6 | 0.257 | 761 |

# Authorship Contribution Statement

**Tugrul Y. Ertugrul:** Conceptualization, Data curation, Formal analysis, Investigation, Methodology, Software, Validation, Visualization, Writing-original draft. **Woong Choi:** Data curation, Investigation, Validation, Writing – review and editing. **Adam Z. Weber:** Supervision, Project administration, Funding acquisition. Resources, Writing – review and editing. **Alexis T. Bell:** Supervision, Project administration, Funding acquisition, Writing – review and editing.

# Declaration of Competing Interest

The authors declare that they have no known competing financial interests or personal relationships that could have appeared to influence the work reported in this paper.

# Acknowledgement

The authors acknowledge Chevron corp. for their financial support.

# Data Availability

The data that support the findings are available on request.

Ethylene

MEA

Continuum modelling

Cost analysis

## Nomenclature

| | |
|---|---|
| $a_w$ | Water activity |
| $a_s$ | Electrochemically active surface area (m$^{-1}$) |
| $c_j$ | Concentration of species *j* (M) |
| $D_j$ | Diffusivity of species *i* (m$^2$ s$^{-1}$) |
| $F$ | Faraday constant (C mol$^{-1}$) |
| $H_{CO_2}$ | Henry's constant for $CO_2$ (mM atm$^{-1}$) |
| $H_0$ | Heaviside step function |
| $i_k$ | Partial current density of each product $k$ (A cm$^{-2}$) |
| $i_{0,k}$ | Exchange current density of product $k$ (A cm$^{-2}$) |
| $i_s$ | Solid phase current density (A cm$^{-2}$) |
| $i_l$ | Liquid phase current density (A cm$^{-2}$) |
| $K_n$ | Equilibrium constant in reaction *n* |
| $K_V$ | Membrane/Ionomer conductivity (S m$^{-1}$) |
| $k_n$ | Forward/backward rate constant of reaction *n* (m$^3$ s$^{-1}$ mol$^{-1}$) |
| $k_{T,m}$ | Thermal conductivity of medium $m$ |
| $k_{MT}$ | Mass transfer coefficient |
| $M_j$ | Molar mass of species *j* (g mol$^{-1}$) |
| $N_j$ | Molar flux of species *j* (mol m$^{-2}$ s$^{-1}$) |
| $n_k$ | Number of electrons transferred in reaction *k* |
| $N_w$ | Molar flux of water in the membrane/ionomer phase (mol m$^{-2}$ s$^{-1}$) |
| $\dot{n}_{C_2H_4}$ | Molar production rate of $C_2H_4$ |
| $P$ | Pressure (Pa) |
| q | Heat transfer (W $m^{-2}$) |
| $Q$ | Heat generation term (W $m^{-3}$) |
| $Q$ | Mass generation/consumption term (kg $m^{-3}$ $s^{-1}$) |
| $r_p$ | Average catalyst particle radius (m) |
| $R$ | Ideal gas constant (J $mol^{-1}$ $K^{-1}$) |
| $R_{B,j}$ | Mole generation/consumption for bulk reactions species *j* (mol $m^{-3}$ $s^{-1}$) |
| $R_{PT,j}$ | Molar rate of phase-transfer for species *j* (mol $m^{-3}$ $s^{-1}$) |
| $U_k^0$ | Standard Thermodynamic potential of product $k$ (V) |
| $S_{CL}$ | Catalyst-layer saturation |
| $T$ | Temperature (K) |
| $u$ | Mass averaged velocity (m $s^{-1}$) |
| $u_{mob,j}$ | Mobility of species *j* ($m^2$ $V^{-1}s^{-1}$) |
| $V$ | Voltage (V) |
| $V_{m,w}$ | Molar volume of water ($m^3$ $mol^{-1}$) |
| $y_j$ | Mole fraction of gaseous species *j* |
| $z_j$ | Charge of ion *j* |

**Greek**

| | |
|---|---|
| $\alpha_{a/c,k}$ | Transfer coefficient of anodic/cathodic reaction *k* |
| $\alpha_w$ | Water transport coefficient (mol $J^{-1}$ $cm^{-1}$ $s^{-1}$) |
| $\gamma_{j,k}$ | Activity coefficient |
| $\epsilon$ | Porosity |
| $\varepsilon_s$ | Solid volume fraction |
| $\varepsilon_l$ | Liquid volume fraction |
| $\zeta_j$ | Electroosmotic drag coefficient of species *j* |
| $\eta_k$ | Overpotential for reaction *k* (V) |
| $\kappa_{m,p}$ | Permeability of phase $p$ and medium $m$ ($m^2$) |
| $\lambda$ | Water content |
| $\mu$ | Viscosity (Pa s) |
| $\mu_w$ | Chemical potential of water ($J\ mol^{-1}$) |
| $\nu_{j,k}$ | Stoichiometric coefficient of species *j* for reaction $k$ |
| $\sigma_s$ | Solid phase electric conductivity ($S\ m^{-1}$) |
| $\Pi_j$ | Peltier coefficient for reaction *j* |
| $\rho$ | Density ($g\ cm^{-3}$) |
| $\phi_s$ | Solid phase potential (V) |
| $\phi_l$ | Liquid phase potential (V) |
| $\psi$ | Saturated permeability |
| $\omega_i$ | Mass fraction of gaseous species *j* |

**Subscript**

| | |
|---|---|
| *CT* | Charge-transfer |
| *eff* | Effective |
| *G* | Gas-phase |
| *I* | Ionomer-phase |
| *j* | Ionic species |
| *k* | Product |
| *l* | Liquid-phase |
| *L* | Liquid-phase |
| *M* | Membrane |
| *PT* | Phase-transfer |
| *MT* | Mass-transfer |
| *p* | Phase |
| *q* | Other gaseous species |
| *s* | Solid phase |
| *sat* | Saturated value |
| *ref* | Reference value |
| *w* | Value in water |

**Superscript**

| | |
|---|---|
| *eff* | Effective |
| *K* | Knudsen |
| *m* | Molecular |
| *0* | Intrinsic value or standard state |
| *ref* | Reference |

**Acronyms**

| | |
|---|---|
| AEM | Anion exchange layer |
| aDL | Anode diffusion layer |
| aCL | Anode catalyst layer |
| aCN | Anode channel |
| cCL | Cathode catalyst layer |
| cDL | Cathode diffusion layer |
| cCN | Cathode channel |
| $CO_2$R | $CO_2$ reduction |
| HER | Hydrogen evolution reaction |
| OER | Oxygen evolution reaction |
| MEA | Membrane electrode assembly |